\documentclass[fleqn,usenatbib]{mnras}

\usepackage[T1]{fontenc}

\DeclareRobustCommand{\VAN}[3]{#2}
\let\VANthebibliography\thebibliography
\def\thebibliography{\DeclareRobustCommand{\VAN}[3]{##3}\VANthebibliography}

\usepackage{graphicx}	
\usepackage{amsmath}	
\usepackage{amssymb}	

\usepackage[]{subfig}
\usepackage{soul}		

\usepackage{xspace}
\usepackage{hyperref}
\usepackage[x11names]{xcolor}
\usepackage{newtxtext,newtxmath}

\newcommand{\magnet}{\mbox{{\sc \small Magneticum}}\xspace}
\newcommand{\boxhr}{\mbox{{\sc \small Box2/hr}}\xspace}
\newcommand{\gadget}{\mbox{{\sc \small GADGET-3}}\xspace}
\newcommand{\msun}{\mbox{$\rm{M}_\odot$}\xspace}

\definecolor{dollarbill}{rgb}{0.72, 0.93, 0.6}
\newcounter{KKcounter}
    
\definecolor{mah}{rgb}{0.85, 0.63, 0.90}
\newcounter{CScounter}

\defcitealias{nusser2000least}{NB00}

\title[Tracing the environmental history of observed galaxies via eFAM reconstruction]
{Tracing the environmental history of observed galaxies \newline via extended fast action minimization method}

\author[E. Sarpa et al.]{
	E. Sarpa,$^{1}$\thanks{E-mail: elena.sarpa@iap.fr}
	A. Longobardi,$^2$
	K. Kraljic,$^3$
    A. Veropalumbo,$^{4,5}$
    C. Schimd$^3$
	\\
$^{1}$ Aix Marseille Univ, CNRS/IN2P3, CPPM, Marseille, France \\
$^{2}$ Dipartimento di Fisica G. Occhialini, Universit\`a degli Studi di Milano Bicocca, Piazza della Scienza 3, 20126 Milano, Italy \\
$^{3}$ Aix Marseille Univ, CNRS, CNES, LAM, Marseille, France \\
$^4$ Dipartimento di Matematica e Fisica, Universit\`a degli studi Roma Tre, Via della Vasca Navale, 84, 00146 Roma, Italy\\
$^{5}$ INFN - Sezione di Roma Tre, via della Vasca Navale 84, I-00146 Roma, Italy\\
}

\date{Accepted XXX. Received YYY; in original form ZZZ}

\pubyear{2015}

\begin{document}
\label{firstpage}
\pagerange{\pageref{firstpage}--\pageref{lastpage}}
\maketitle

\begin{abstract}

We present a novel application of the extended Fast Action Minimization method (eFAM) aimed at assessing the role of the environment in shaping galaxy evolution and validate our approach against the Magneticum hydrodynamical simulation. We consider the $z\simeq 0$ snapshot as our observed catalogue and use the reconstructed trajectories of galaxies to model the evolution of cosmic structures. At the statistical level, the fraction of volume occupied by voids, sheets, filaments, and clusters in the reconstructed and simulated high-redshift snapshots agree within $1\sigma$. Locally, we estimate the accuracy of eFAM structures by computing their purity with respect to simulated structures, $P$, at the cells of a regular grid. Up to $z=1.2$, clusters have $0.58<P<0.93$, filaments vary in $0.90<P<0.99$, sheets show $0.78<P<0.92$, and voids have $0.90<P<0.92$. 
As redshift increases, comparing reconstructed and simulated tracers becomes more difficult and the purity decreases to $P\sim 0.6$.
We retrieve the environmental history of individual galaxies by tracing their trajectories through the cosmic web and relate their observed gas fraction, $f_\mathrm{gas}$, with the time spent within different structures. 
For galaxies in clusters and filaments, eFAM reproduces the dependence of $f_\mathrm{gas}$ on the redshift of accretion/infall as traced by the simulations with a 1.5$\sigma$ statistical agreement (which decreases to 2.5$\sigma$ for low-mass galaxies in filaments). These results support the application of eFAM to observational data to study the environmental dependence of galaxy properties, offering a complementary approach to that based on light-cone observations.

\end{abstract}




\section{Introduction}
In the cold dark matter scenario, the observed Universe originates from the gravitational amplification of primordial density perturbations. On small scales (below a few Mpc), dark matter haloes arise from the collapse of dark matter perturbations and gradually grow in size by accretion and merging with nearby companions. Later on, baryons cool and collapse to the depths of the gravitational potential forming galaxies within these haloes \citep{White1978MNRAS.183..341W}. On a larger scale ($\geq 10$ Mpc), haloes and galaxies move along the gradient of the gravitational potential dressing the large-scale structure detected in systematic galaxy redshift surveys \citep[e.g.][]{DeLapp1986ApJ...302L...1D,Geller1989Sci...246..897G,2df2001MNRAS.328.1039C,sdss2004ApJ...606..702T}. 
The emerging web-like pattern, dubbed cosmic web \citep{ Bond1996Natur.380..603B}, is a compound of large voids surrounded by sheet-like walls and filamentary structures connecting collapsing galaxy clusters.
From this perspective, the formation and evolution of galaxies and cosmic web are expected to be closely intertwined. As galaxies move within the cosmic web, they evolve by experiencing the gravitational and hydrodynamical influence proper to their hosting environments.

The imprint of the large-scale environment on the observed galaxy properties has been recently established by the analysis of wide spectroscopic surveys such as the Sloan Digital Sky Survey \citep[SDSS; SDSS IV MaNGA,
][]{Alam2015ApJS..219...12A, Bundy2015,Spindler2018}, VIMOS Public Extragalactic Redshift Survey \citep[VIPERS,][]{Davidzon2016,Cucciati2017, Malavasi2017,Scodeggio2018A&A...609A..84S}, the Cosmic Evolution Survey \citep[COSMOS,][]{Scoville2013,COSMOS20152016ApJS..224...24L,Laigle2018MNRAS.474.5437L,Betti2019}, and Galaxy and Mass Assembly survey \citep[GAMA,][]{GAMA2011MNRAS.413..971D,Kraljic2018MNRAS.474..547K}. However, it is still debated whether the environmental dependence arises at the time of galaxy formation within over-dense regions (nature scenario) or whether it is the time-integrated effect of gravitational and hydrodynamical interactions with the environment (nurture hypothesis). For a review on environmental effects on cluster galaxies see \cite{Boselli2006PASP..118..517B,Boselli2021arXiv210913614B}.

To date, the efforts on disentangling the nature vs nurture hypothesis follow two complementary approaches: either statistically inferring the evolution of galaxy properties by observing galaxies populations across a wide redshift range \citep[e.g.,][]{Ilbert2013, Cucciati2017, Malavasi2017}, or using cosmological simulations to relate late-time galaxy properties to their individual environmental history \citep[e.g.,][]{DeLucia2012,Gheller2016,Donnari2021}. The observational approach provides direct evidence of galaxy evolution. However, its accuracy is hampered by the poor statistics of the observed samples at high redshifts and can ultimately lead to conflicting results (e.g. \citealp{Guo2015} and \citealp{Eardley2015}).
On the other hand, modern cosmological simulations reproduce a large range of observational results, however, they adopt phenomenological models of many of galaxy formation processes, details of which remain poorly understood.

A valid alternative to the methods mentioned above is offered by back-in-time reconstruction techniques, which are able to recover both the dynamics and the initial conditions of an observed galaxy sample. Since their first application to trace galaxy orbits in the Local Group \citep{Peebles1989ApJ...344L..53P}, these techniques have been successfully employed to study the dynamics of the observed Universe \citep{Branchini2002,Romano2005A&A...440..425R,Shaya2017}, with particular emphasis on the modelling of the galaxies peculiar velocity field \citep{Mohayaee2004,Kitaura2012,Graziani2019}. Thanks to the development of forward-modelling and machine learning techniques, reconstruction is entering a new era, establishing itself as a complementary tool to investigate the formation and evolution of the cosmic web \citep{Leclercq2015JCAP...06..015L,Kitaura2021}. 

In this work, we propose to employ the extended Fast Action Minimization reconstruction method \citep[eFAM;][]{eFAM12019MNRAS.484.3818S,eFAM22021MNRAS.tmp..398S}, initially designed for baryon acoustic oscillation studies, to extend the deterministic approach proper to simulations for the analysis of observed properties of galaxies. By applying eFAM to a simulated sample of galaxies, we aim at retrieving their past, non-linear trajectories (position and velocity as a function of time) and thus infer the evolution of the large-scale cosmic web.
We then combine this information to extract the environmental history of individual objects and investigate the dependence of the gas fraction in cluster and filament galaxies measured at the redshift of the observations on their redshift of infall into the structure.

The paper is structured as follows. In Section~\ref{sec:data}, we describe the set of simulated catalogues used in this study, in Section~\ref{sec:methods:eFAM} we overview the extended Fast Action Minimization method, and we detail the fiducial metric for the description of the cosmic web in Section~\ref{sec:methods:Tweb}. We dedicate Section~\ref{sec:analysis} to the validation of eFAM results against simulations, particularly focusing on \emph{i}) assessing the quality of our cosmic web reconstruction (Section~\ref{sec:reconCW}) and \emph{ii}) verifying the accuracy of the retrieved trends in the evolution of the gas fraction within clusters and filaments (Section~\ref{sec:Env_gal}). We discuss our results in Section~\ref{sec:discussion} and present our summary and conclusions in Section~\ref{sec:summary}.

\vspace{-2mm}
\section{Data: Magneticum simulations}\label{sec:data}

We test the eFAM reconstruction algorithm on the cosmological hydrodynamical simulation Magneticum Pathfinder\footnote{\href{http://www.magneticum.org}{http://www.magneticum.org}}, hereafter \magnet, a suite of cosmological simulations run with the
smoothed particle hydrodynamics solver \gadget \citep{Springel2005}
and with fiducial cosmology compatible with WMAP7 \citep{WMAP72011ApJS..192...18K} $\Lambda$CDM cosmology ($\Omega_\mathrm{m}=0.272$, $\Omega_\Lambda=0.728$, $\Omega_\mathrm{b}=0.0456$ $h=0.704$, $n_\mathrm{s}=0.963$ and $\sigma_8=0.809$).
We employ the \boxhr simulation with 352 $h^{-1}\mathrm{Mpc}$ box length and an initial total particle number of $2 \times 1584^3$ corresponding to a mass resolution for the dark matter (DM), gas, and stellar particles of
$m_{\rm DM} = 1.4\times10^9 h^{-1}\msun$, $m_{\rm gas} = 2.7\times10^8 h^{-1}\msun$, and $M_{\star} = 7.1\times10^7{h^{-1} \msun}$, respectively. The fiducial values for the softening parameter are $\epsilon_{\rm DM}=\epsilon_{\rm gas}= 3.5$~$h^{-1}$kpc for DM and gas particles, and $\epsilon_{\star}=2$~$h^{-1}$kpc for stellar particles.  
We refer to \cite{Mag12014MNRAS.442.2304H}, \cite{Tekluetal2015} and \cite{Mag22017A&C....20...52R} for a detailed description of the simulation, focusing here only on the features of interest for the present analysis.  
The simulation includes radiative cooling \citep{Wiersma2009}, heating from a uniform time-dependent ultraviolet/X-ray background \citep{Haardt_Madau2001}, star formation based on the multiphase model of the interstellar medium \citep{Springel2003}, and the associated feedback processes. These include the explosions of Type II supernovae (SNII) triggering galactic winds with a velocity of 350 km s$^{-1}$. Chemical evolution is modelled according to \cite{Tornatore2007} following the metal production by SNII, supernovae of Type Ia, and asymptotic giant branch stars.
The growth and dynamics of black holes and associated feedback from active galactic nuclei in \magnet
are based on the model of \cite{Springeletal2005} and \cite{DiMatteoetal2005}, with the modifications as described in \cite{Mag12014MNRAS.442.2304H}.

In the analysis presented here, we treat the snapshot at $z=0.07$ as the observed catalogue, referring to it as $z_\mathrm{obs}$, and consider only the simulated properties of galaxies that are available in real datasets, i.e. the galaxies stellar mass and gas content at $z_\mathrm{obs}$, and their observed positions in redshift-space. To match as closely as possible the galaxy number density of low-redshift spectroscopic galaxy surveys such as GAMA \citep{Baldry2012,delaTorre2013} and VIPERS \citep{delaTorre2013}, both of about $10^{-2}h^3\mathrm{Mpc}^{-3}$, we consider solely the galaxies with stellar mass above $10^9h^{-1}\mathrm{M}_{\sun}$.
We use the snapshots at $z$ = 0.13, 0.22, 0.30, 0.37, 0.47, 0.68, 0.79, 0.90, 1.03, 1.33, 1.71, 2.33 as a benchmark for the reconstructed evolution of cosmic web and galaxy properties.

\vspace{-2mm}
\section{Methods}

\subsection{Galaxy orbits reconstruction: eFAM}
\label{sec:methods:eFAM}

To assess the environmental history of observed galaxies, we must have access to both their past trajectories and the evolution of the large-scale structure. In this work, we extract this information from the observed position of galaxies using the extended Fast Action Minimization method
\citep[eFAM;][]{eFAM12019MNRAS.484.3818S,eFAM22021MNRAS.tmp..398S}. eFAM is a back-in-time reconstruction technique based on the Least Action Principle \citep{Peebles1989ApJ...344L..53P} apt at recovering the trajectories of the mass tracers, i.e. position and velocity as a function of time, by minimizing the action of the system under mixed boundary conditions. Assuming that galaxies trace the underlying mass distribution, boundary conditions are the observed positions of galaxies in redshift-space, $\mathbf{s}_{i,\mathrm{obs}},$ and their initial peculiar velocities, which we assume to vanish at early times in agreement with the cosmological principle.

In the eFAM framework, galaxies are identified with their centre-of-mass and treated as collisionless, equal mass, point-like particles interacting only by gravity in an expanding universe. In the weak-field limit, the action of a system of $N$ galaxies with comoving coordinates $\{\mathbf{x}_i\}_{i=1,...N}$ is 
\begin{eqnarray}\label{eq:action}
S&=&\sum_{i=1}^N\int_0^{D_\mathrm{obs}} dD\left[ fEDa^2\frac{1}{2}\left(\frac{d\mathbf{x}_i}{dD}\right)^2+\frac{1}{2fDE}\phi(\mathbf{x}_i,D)\right]\nonumber \\
&&+\frac{1}{2}\sum_{i=1}^N\left(fEDa \frac{d\mathbf{x}^\parallel_i}{dD}\right)_\mathrm{obs}^2,
\end{eqnarray}
where $D$ is the linear growth factor, used as time variable, $f=d\ln D/d\ln a$ the growth rate, $E = H/H_0$ the dimensionless Hubble parameter, and $a$ the scale factor. The first term in the action is the kinetic energy. The second term describes the potential energy depending on the peculiar gravitational potential $\phi(\mathbf{x})$ solving the Poisson equation 
\begin{equation}\label{eq:Poisson}
\nabla^2\phi(\mathbf{x})=4\pi Ga^2\bar{\rho}\delta_\mathrm{g}(\mathbf{x})/b.
\end{equation}
Here, $G$ is the gravitational constant, $\bar{\rho}$ the mean density of the Universe, and $\delta_\mathrm{g}=n_\mathrm{g}/\bar{n}_\mathrm{g}-1$ is the number density fluctuation of galaxies with respect to the average number density, $\bar{n}_\mathrm{g}$. As in \cite{eFAM12019MNRAS.484.3818S,eFAM22021MNRAS.tmp..398S}, $\phi(\mathbf{x})$ is computed on the unsmoothed galaxy distribution using the falcON \citep{FalcOn2002} algorithm, an optimised Poisson solver combining the tree-code and fast
multipole methods.
The linear galaxy bias, $b$, allows us to model the effective gravitational potential due to both the and dark matter distributions by relating $\delta_\mathrm{g}$ to the total matter density fluctuation, $\delta=\rho/\bar{\rho}-1$, via $\delta=\delta_\mathrm{g}/b$. In our reconstruction, we fix $b$ to the value obtained by fitting the two-point correlation function, $\xi$, of the galaxy distribution at the observed redshift with the fiducial template \citep{Marulli2017} 
\begin{equation}\label{eq:fit_b}
    \xi(s)=\left[(b\sigma_8)^2+\frac{2}{3}fb\sigma_8^2+\frac{2}{5}(f\sigma_8)^2 \right]\frac{\xi_\mathrm{DM}(r)}{\sigma^2_8}.
\end{equation}
Here, $\xi_\mathrm{DM}$ is the expected dark matter correlation function at the desired redshift estimated via the {\small CAMB} software \citep{Lewis2002} and $\sigma_8$ is the amplitude of the (linear) power spectrum on a scale of $8 h^{-1}\mathrm{Mpc}$. By fixing $b$ to its observed value, we consider galaxies as if they were tracing peaks in the underlying total matter density field with an amplitude proportional to $(1+\delta)/(1+\delta_\mathrm{g})$  \citep[see Section 4.3 in ][]{NusserBranchini2000}, effectively reconstructing the trajectories of the centre of mass of these overdensities.

The unobserved matter distribution beyond the survey volume is assumed to be uniform and does not contribute to the action. 
The third term of the action is the correction needed to enable the minimization in redshift-space. 
Finally, the superscript $^\parallel$ marks the projection of the position vector along the galaxy's line-of-sight, and the subscript $_\mathrm{obs}$ refers to quantities computed at the redshift of observations, $z_\mathrm{obs}$. 

The reconstruction is performed by modelling the orbits of galaxies as a linear combination of $M$ time-dependent basis functions ${q_n(D)}_{n=1...,M}$ viz.
\begin{equation}\label{eq:traj}
\mathbf{x}_i(D)=\mathbf{x}_{i,\mathrm{obs}}(D)+\sum_{n=1}^M\mathbf{C}_{i,n}q_n(D),
\end{equation}
and searching for the set of expansion coefficients $\mathbf{C}_{i,n}$ that minimize the action.

As pointed out by \cite{Branchini2002, eFAM12019MNRAS.484.3818S,eFAM22021MNRAS.tmp..398S}, the solution of the least action problem is in general not unique. To circumvent this issue, we bound the solution to the global minimum of the action by using the Zel’dovich approximation \citep{zel'dovic1970} as the first guess in the action minimisation. This approach does not constrain the reconstructed trajectories to the linear regime allowing us to fairly model the ``true'' $N$-body peculiar velocities of the observed galaxies outside the virialized core of structures. The treatment of virialized structures is discussed in Section~\ref{sec:methods:FoG}.

Before presenting eFAM results, we shall briefly discuss the physical implication of its underlying assumptions. Similarly to many reconstruction algorithms \citep[e.g.][]{NAM,ZArec}, eFAM conserves the number of mass tracers thus failing to recover the merging history of the traced dark matter halos. In addition, mass tracers and host halos are assumed to follow the same trajectory. This approximations do not affect the reconstructed large-scale matter clustering \citep{eFAM22021MNRAS.tmp..398S}, however, it might lower the accuracy of individual trajectories at $z>1.5$, where major mergers are frequent. When analysing individual trajectories in Section~\ref{sec:gas_fraction} and \ref{sec:gas_fraction_fl}, we will therefore focus on the redshift range $0.07\leq z \leq 1.2$.
A second eFAM approximation is to overlook the role of galaxy mass loss and accretion in shaping the trajectory of the traced matter overdensity. We justify this assumption by noting that the associated mass variation is negligible compared to the mass of the hosting halo. 
In Appendix~\ref{app:velocities}, we test the robustness of our assumptions by characterising eFAM performances in different environments and comparing the results with the Zel'dovich solution. We conclude that eFAM fairly models the simulated peculiar velocities of observed galaxies in different environments, outperforming Zel'dovich reconstruction outside virialized regions.  Within the core of clusters, the thermal motion of galaxies prevents both eFAM and ZA to recover the ``true'' orientation of the velocity field. Nonetheless, eFAM is superior to ZA in modelling the correct velocity amplitude. 


\vspace{-2mm}
\subsection{Fingers-of-God: correction by friend-of-friend}
\label{sec:methods:FoG}
High-resolution galaxy catalogues, i.e. with mean galaxy number density $\bar{n} \approx 10^{-2}h^3\mathrm{Mpc}^{-3}$, allow us to resolve small scale structures $(\lesssim 1 h^{-1}\mathrm{Mpc})$ providing exceptional insights into the dynamics of these systems on a wide range of scales. Yet, when observed in redshift-space, the large velocity dispersion of virialized regions, such as groups and clusters of galaxies, cause structures to appear smeared along the line-of-sight forming the so called Fingers-of-God, impacting the accuracy of the cosmic web classification. Reconstruction techniques like eFAM can compensate for the redshift-space distortions (RSD) by recovering the velocities of galaxies and, consequently, their real-space positions. Nonetheless, as mentioned above, the current implementation of eFAM is not optimized to model RSD in the very interior of groups and clusters. 
To circumvent this issue, we identify galaxy groups and cluster cores by running the \textsf{pyfof}\footnote{https://pypi.python.org/pypi/pyfof} friend-of-friend cluster finder on the real-space galaxy catalogue at $z_\mathrm{obs}=0.07$ and collapsing each group into point-like objects. To each collapsed particle, we then assign a mass corresponding to the sum of the masses of the group members and a peculiar velocity equal to the one of their centre-of-mass. To conserve the original local number density, we weight compressed objects by the number of its group members. 

The group classification depends on the fiducial linking length, $l_\mathrm{g}$, used to identify group members. Here, we set $l_\mathrm{g}$ so as to minimise the error in the inferred real-space galaxy position. We run 100 reconstructions varying $l_\mathrm{g}$ between 0.01 and $1 h^{-1}\mathrm{Mpc}$ and find that a value of $l_\mathrm{g}=0.3 h^{-1}\mathrm{Mpc}$ reduces the mean error on the real-space positions from $2.16 h^{-1}\mathrm{Mpc}$ prior to RSD correction, down to $1.3 h^{-1}\mathrm{Mpc}$ after accounting for eFAM velocities. With this choice, we restrain the compression to small groups of galaxies and the virialized core of clusters ($\lesssim20$ objects), effectively grouping the 72 per cent of satellite galaxies within the mass range $9<\log_{10}(M_\star/h^{-1}M_\odot)<12$ and the 26 per cent of all galaxies. From the pre-reconstruction sample, we notice that groups with less than five members do not significantly contribute to Fingers-of-God; therefore, we decide to limit the compression to larger groups.
This procedure can be extended to redshift-space by allowing two different linking-lengths for the pair separation transverse to and along the line-of-sight \citep[see e.g.][]{sfof2011MNRAS.417.1402F,amico2018MNRAS.473.5221B}
\begin{figure*}
\centering
\includegraphics[width=1.\textwidth]{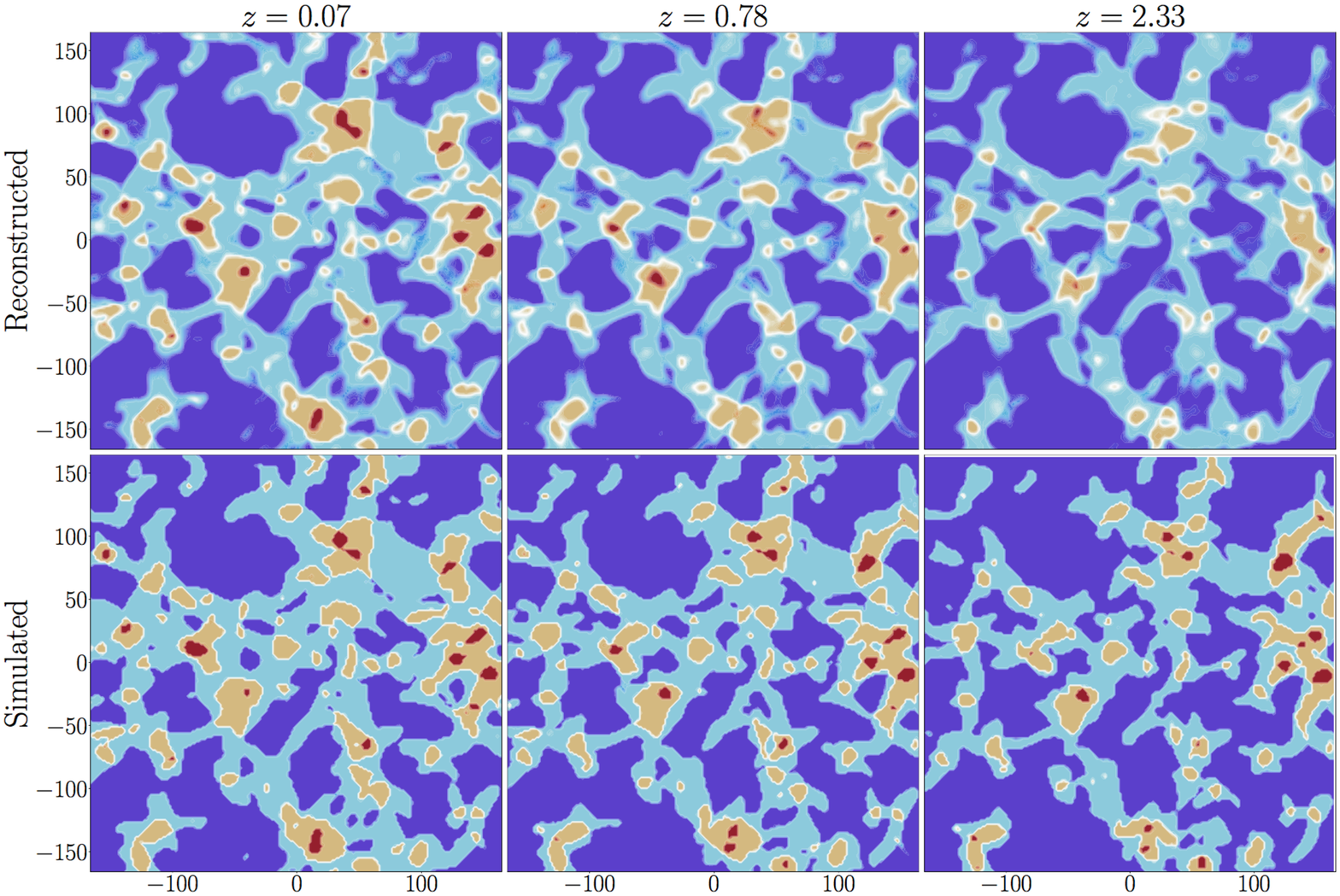}
\caption{Visual comparison of cosmic web patterns as detected in the reconstructed (top row) and simulated (bottom row) galaxy catalogues at three different redshifts. Each panel illustrates the segmentation of structures in a $1.68 h^{-1}\mathrm{Mpc}$ thick slice. Different colours correspond to voids (dark-blue), sheets (light-blue), filaments (yellow), and clusters (red). Units are in $h^{-1}$Mpc.}
\label{fig:C-web_Visual}
\end{figure*}

\vspace{-2mm}
\subsection{Reconstructed catalogues}
After removing very-low mass galaxies and compressing the Fingers-of-God, we run eFAM on the ``observed'' galaxy catalogue in redshift-space. To better capture the non-linear motion within high-density regions, we allow the polynomial expansions in Equation~\ref{eq:traj} to reach the 10-th order, i.e. the maximum order allowed by our current computational resources. Finally, we use the reconstructed trajectories of the mass tracers to remove RSD from the observed galaxy distribution and build a series of higher redshift catalogues describing the tracers positions at the redshifts of the snapshots of the simulation considered in this work (see Section~\ref{sec:data} for details on the catalogues). 

\vspace{-2mm}
\subsection{Cosmic web classification: T-web}
\label{sec:methods:Tweb}

Among the various cosmic web classifiers \citep[see][for a comparative review]{Libeskind2018}, we choose to apply to each simulated and reconstructed catalogues the T-web classification scheme proposed by \cite{Hahn2007MNRAS.375..489H}. The T-web specifies morphological components locally at the cells of a cubic grid covering the simulated volume, and large-scale structures are three-dimensional bodies identified by the set of adjacent cells of the same type. Tracing galaxy orbits within these three-dimensional structures allows us to determine their environmental history. The classification procedure can be summarized as follows:
\emph {i)} for each sample of mass tracers we estimate the linear bias $b(z)$ by fitting Equation~\eqref{eq:model} to the measured two-point correlation function of matter tracers, i.e. galaxies when dealing with simulations and the centre-of-mass of matter overdensities when dealing with reconstructed catalogues. We verify that $b$ increases with redshift in the case of simulated galaxies while remaining unchanged for reconstructed tracers. 
\emph{ii)} We estimate the number density field, $n(\mathbf{x})$, by performing a clouds-in-cell interpolation on an equally-spaced cubic grid of step $1.68 h^{-1}\mathrm{Mpc}$ covering the simulation box and smoothing the interpolated field with a Gaussian kernel of size $R_\mathrm{s}$ (see below).
\emph{iii)} At each grid point, we compute the peculiar gravitational potential $\phi(\mathbf{x},z)$ solving Equation~(\ref{eq:Poisson}). Then, we \emph{iv)} evaluate the eigenvalues $\mu_1\geq\mu_2\geq\mu_3$ of the tidal field $\partial_i\partial_j\phi(\mathbf{x})$, with $\partial_i$ denoting the spatial partial derivative along the $i$-th Cartesian direction ($i,j=1,2,3$) and \emph{v)} classify cells according to $\{\mu_j\}_{j=1,2,3}$. Considering as positive the eigenvalues beyond the fiducial threshold $\mu_\mathrm{th}$, three positive eigenvalues mark a cell of type cluster, two positive eigenvalues identify a filaments, and one and zero positive eigenvalues label a sheet and a void, respectively.

It is worth noting that by using a cosmic web classification based on the profile of the gravitational potential, we label as a cluster the whole potential well of the central overdensity, effectively including a region a few times larger than its virial radius.


The choice of $R_\mathrm{s}$ strongly affects the local eigenstructure of the gravitational potential by varying the typical length-scale of structures identified as stable clusters \citep[see Section 3.3 in][]{Hahn2007MNRAS.375..489H}. Here, we fix $R_\mathrm{s}$ to the radius of a sphere with uniform density $\bar{\rho}$ and total mass, $M_\mathrm{s} = (2\pi)^{3/2}\bar{\rho}R_\mathrm{s}^3$, matching the minimum mass of our target clusters.
Since we are interested in massive clusters with total mass above $10^{14} h^{-1}M_\odot$,
we set $R_\mathrm{s}=4.38 h^{-1}\mathrm{Mpc}$.
Besides, the T-web classification depends on the fiducial value of the threshold $\mu_\mathrm{th}$ defining collapsing structures.
A natural choice is to fix $\mu_\mathrm{th}=0$ as prescribed in \citet{zel'dovic1970}. We test the robustness of the corresponding classification with respect to Poisson-noise in the sampled density field by running the T-web finder on a uniform distribution of point-like synthetic objects sampling the simulation volume with the same number density as our observed galaxy catalogue. We find that $\mu_\mathrm{th}=0$ yields 89 per cent of cells classified as collapsing structures, in contrast with the un-clustered nature of the synthetic sample. This result is in agreement with the analysis presented in \cite{Forero2009MNRAS.396.1815F}, and shows that the Zel'dovich prescription overestimates the fraction of collapsing structures in disfavour of voids, hence failing to capture the visual impression of the cosmic web.

To obtain a more robust classification of structures, we search for the optimal $\mu_\mathrm{th}$ by performing 1000 T-web classifications of the synthetic catalogue with $0<\mu_\mathrm{th}<2$. For every classification, we compute the fraction of the simulated volume occupied by different structures and fix the fiducial value of $\mu_\mathrm{th}$ to the one yielding a unitary fraction of voids. 
For a synthetic catalogue with the same number density as the observed catalogue, we find $\mu_\mathrm{th}=1.3$. 
High-redshift reconstructed catalogues share the same number density as the observed catalogue. Thus, we can fix $\mu_\mathrm{th}(z)$ to its fiducial value at the observed redshift, $z_\mathrm{obs}$.
Conversely, the number density of simulated galaxies steeply decreases for $z>0.5$. We account for the different amplitudes of the Poisson noise by repeating the procedure described above at each redshift and consistently varying the number density of the synthetic catalogue to match that of the snapshots.

\vspace{-2mm}
\section{Analysis}\label{sec:analysis}

\subsection{Cosmic web reconstruction}
\label{sec:reconCW}
We devote this section to the analysis of the reconstructed cosmic web as a function of redshifts, presenting our results in increasing order of detail. In Section~\ref{subsection:Visual} we provide a visual comparison between the simulated and the reconstructed webs, in Section~\ref{subsection:VFF} we quantify their agreement in terms of volume partitioning between different structure types, and we conclude in Section~\ref{subsection:Purity} by measuring the purity of the reconstructed structures.
\vspace{-2mm}
\subsubsection{Visual comparison}\label{subsection:Visual}
Figure~\ref{fig:C-web_Visual} compares $1.68 h^{-1}\mathrm{Mpc}$ thick slices of the cosmic patterns identified on three reconstructed catalogues at different redshifts (top panels) with the ones estimated from the corresponding simulated snapshots (bottom panels). The reconstructed and simulated webs show a qualitatively similar trend in terms of both volume partitioning and redshift evolution. At all redshifts, voids (dark-blue) and sheets (light-blue) occupy the majority of the volume, followed by filaments (yellow) and clusters (red). As the redshift decreases and the density distribution departs from homogeneity (right to left), the volume fraction occupied by sheets and filaments increases. Simultaneously, cluster regions grow in size while preserving the position of the centre-of-mass of their progenitors. At
$z=2.33$, simulated clusters do not find correspondence in the reconstructed catalogue;
this discrepancy is a consequence of the compression of virialized structures carried out prior to reconstruction (see Section~\ref{sec:methods:FoG}), which forces the core of clusters to be represented by single points at all redshifts, thus compressing their potential well at higher $z$. As shown in Section~\ref{sec:accretion}, this particular behaviour of reconstructed clusters does not impact the study of cluster galaxies properties since it occurs at $z>1$ where only a few galaxies belong to clusters. Finally, Figure~\ref{fig:C-web_Visual}, clearly shows that the reconstructed cosmic web fairly reproduces the spatial distribution of the different structures in the simulated field.

\vspace{-2mm}
\subsubsection{Volume filling fraction}\label{subsection:VFF}

\begin{figure}
\centering
\includegraphics[width=\columnwidth]{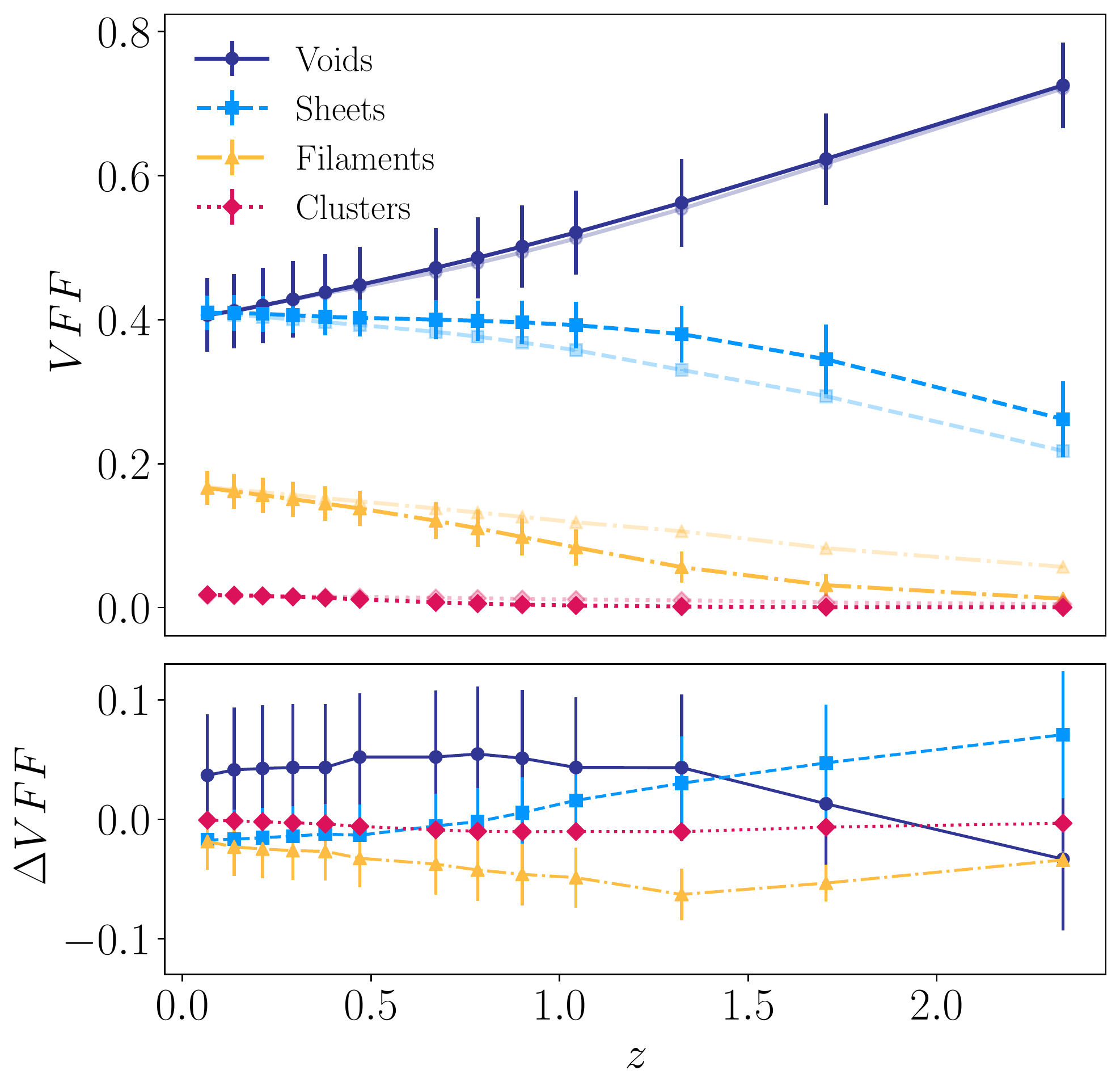}
\caption{\emph{Top panel:} volume filling fraction 
of voids (blue/solid), sheets (light-blue/dashed), filaments (yellow/dot-dashed), and cluster (red/dotted) cells as a function of redshift in reconstructed fields (thick lines) and simulation (thin lines). Error bars represent the standard deviations obtained by rescaling the eigenvalues in the T-web classification by 100 values of the bias drawn from a Gaussian distribution with mean and variance set by the galaxy two-point correlation function modelled according to Equation~\eqref{eq:fit_b}. \emph{Bottom panel:} residuals between reconstructed and simulated volume filling fraction as a function of redshift.  }\label{fig:VFF}
\end{figure}

We quantify the global agreement between the reconstructed and simulated cosmic webs by means of the Volume Filling Fraction (VFF) of different structures types. For grid-based classifiers, the VFF is defined as $VFF_i=N_\mathrm{i}/ N$, where the index $i=\mathrm{V,S,F,C}$ refers to voids, sheets, filaments, and clusters, respectively, $N_i$ is the number of cells of type $i$, and $N$ is the total number of cells covering the volume. Having imposed that a homogeneous distribution is described solely by voids, we expect $VFF_\mathrm{V}$ to increase with redshift and $VFF_\mathrm{S}$, $VFF_\mathrm{F}$, and $VFF_\mathrm{C}$ to decrease.

The top panel of Figure~\ref{fig:VFF} shows the evolution of the reconstructed (thick lines) and simulated (shaded lines) VFFs as a function of redshift. The colour code is the same as in Figure~\ref{fig:C-web_Visual}. 
At each redshift, error bars represent the standard deviations obtained by rescaling the eigenvalues in the T-web classification by 100 values of the bias drawn from a Gaussian distribution with mean and variance set by modelling the galaxy two-point correlation function (see Equation~\ref{eq:fit_b}). The evolution of the VFF confirms the results presented in Figure~\ref{fig:C-web_Visual}. In both datasets, voids and sheets dominate the web, with the former increasingly permeating the space as the distribution becomes more homogeneous. Simultaneously, sheets, filaments, and clusters occupy a lower fraction of volume with increasing redshift. The reconstructed and simulated VFF agree within $1\sigma$ across the redshift range $0.07<z<1.2$ as shown in the bottom panel of Figure~\ref{fig:VFF} where we plot the residuals between the two curves. At $z>1.2$, eFAM reconstructions underestimates the volume occupied by filaments in favour of sheets.

\vspace{-2mm}
\subsubsection{Purity of structures}\label{subsection:Purity}
\begin{figure}
\centering
\includegraphics[width=\columnwidth]{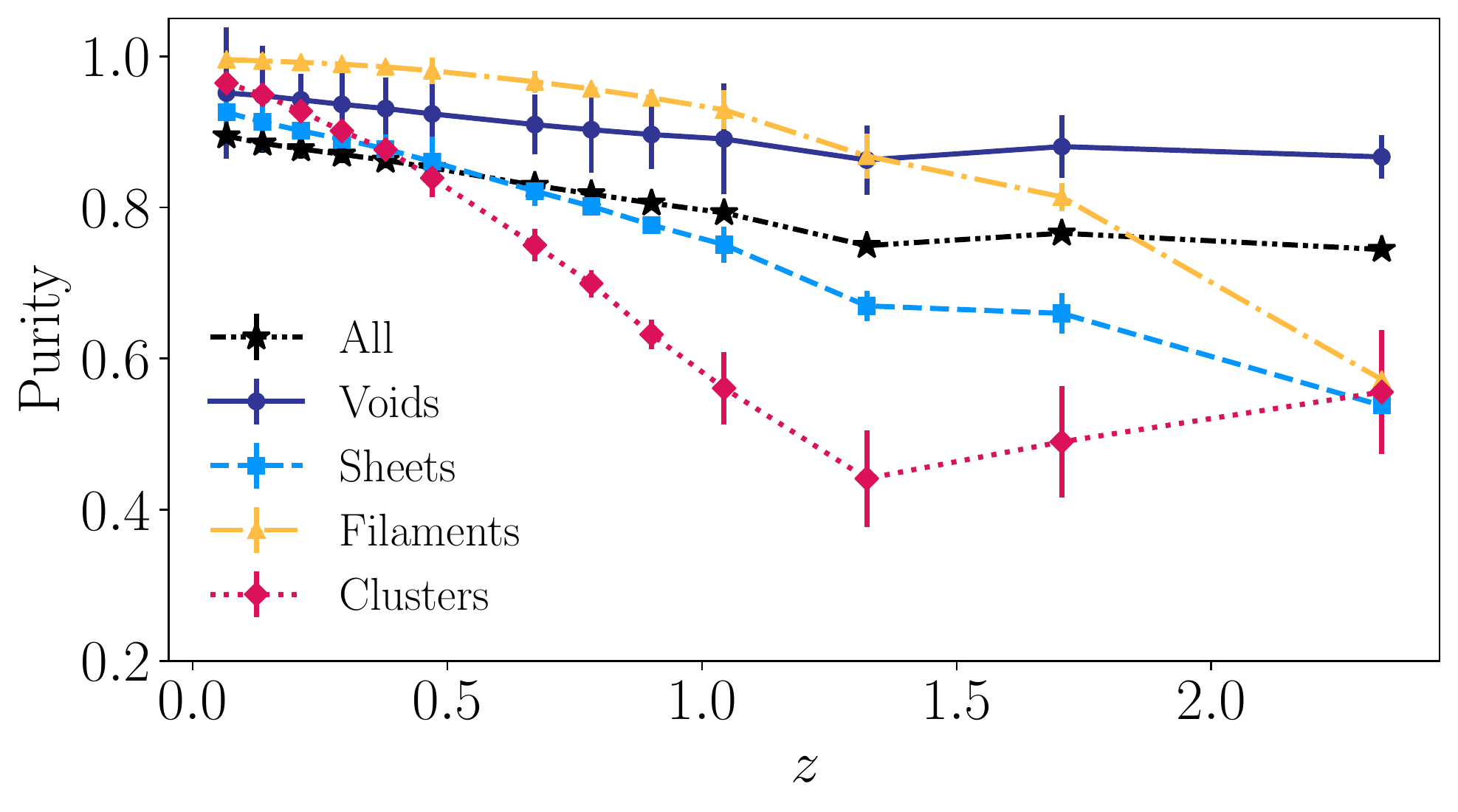}
\caption{Purity of reconstructed structures as a function of redshift. Colour code as in Figure~\ref{fig:VFF}. Error bars are standard deviations computed as in Figure~\ref{fig:VFF}.}\label{fig:Purity}
\end{figure}

The standard approach to assess the quality of a structure finder is to measure the purity of the identified structures with respect to a reference catalogue \citep[e.g.][]{sfof2011MNRAS.417.1402F,Laigle2018MNRAS.474.5437L,Sarron2019}. Here, we quantify the local accuracy of the reconstructed cosmic web at different redshifts by direct comparison with the simulated snapshots.
For each redshift, we select all the grid cells of type $i=\{\mathrm{V},\mathrm{S},\mathrm{F},\mathrm{C}\}$ in the reconstructed catalogue and compute the Euclidean distance between each cell and its nearest neighbour of the same kind in the simulated volume. Then, we count as matched all the cells whose nearest neighbour is closer than a fixed distance $d_i$. Finally, we estimate the purity of the cells of type $i$ as $P_i=N_{i,\mathrm{match}}/N_i$, where $N_{i,\mathrm{match}}$ is the number of matches and $N_i$ is the number of cells of type $i$ in the reconstructed catalogue. $P_i=1$ marks a perfect identification while $P_i=0$ indicates no superposition between the catalogues. For voids and sheets, which occupy most of the volume, we require for each cell to conserve its type in the two datasets, i.e. $d_i=0$. For filaments and clusters we set $d_i=2.9$, corresponding to the diagonal of the cubic cell, thus allowing matching between adjacent cells.

Figure~\ref{fig:Purity} shows the evolution of $P_i$ as a function of redshift. Error bars are estimated following the same procedure adopted for the VFF. At all redshifts and when no distinction between types is considered (black double-dot-dashed line), eFAM correctly identifies more than 80 per cent of cells. Voids are best classified with purity varying between $P_\mathrm{V}=0.95$ at $z_\mathrm{obs}$ and $P_\mathrm{V}=0.87$ at $z=2.3$. Filaments have purity ranging from $P_\mathrm{F}=0.99$ to $P_\mathrm{F}=0.57$,
and sheets follow with purity decreasing from $P_\mathrm{S}\gtrsim0.93$ to $P_\mathrm{S}\simeq0.50$ in the same redshift range.
This trend reflects the difficulty of comparing reconstructed tracers and high redshift galaxies when their number densities and biases are very different. Indeed, while eFAM preserves the number density of the mass tracers by recovering the past position of their host dark matter haloes, \magnet provides the catalogues of existing galaxies at the redshift of the snapshot. 
At $z>1$, the reconstructed clusters lower their purity from $P_\mathrm{C}=0.95$ to $P_\mathrm{C}\simeq0.50$. This result is due to the compression of the virialized structures performed before reconstruction, which prevents tracing the accretion of the innermost cluster galaxies, and confirms that clusters are the most difficult structure to reconstruct. A similar conclusion can be drawn in an indirect way by comparing the two-point correlation functions of simulated and reconstructed tracers at different redshift (see Appendix~\ref{app:xi_comparison}).
In Section~\ref{sec:accretion}, we show that the misclassification of clusters at high redshifts does not affect the quality of the reconstructed environmental history of galaxies since less than 5 per cent of cluster members are accreated before $z=1$.

This analysis shows that filaments are better identified than sheets despite their higher degree of non-linearity. We attribute this result to our T-web finder whose sheets classification is the most sensitive to variations in the fiducial bias.

\vspace{-2mm}

\begin{figure}
\centering
\includegraphics[width=\columnwidth]{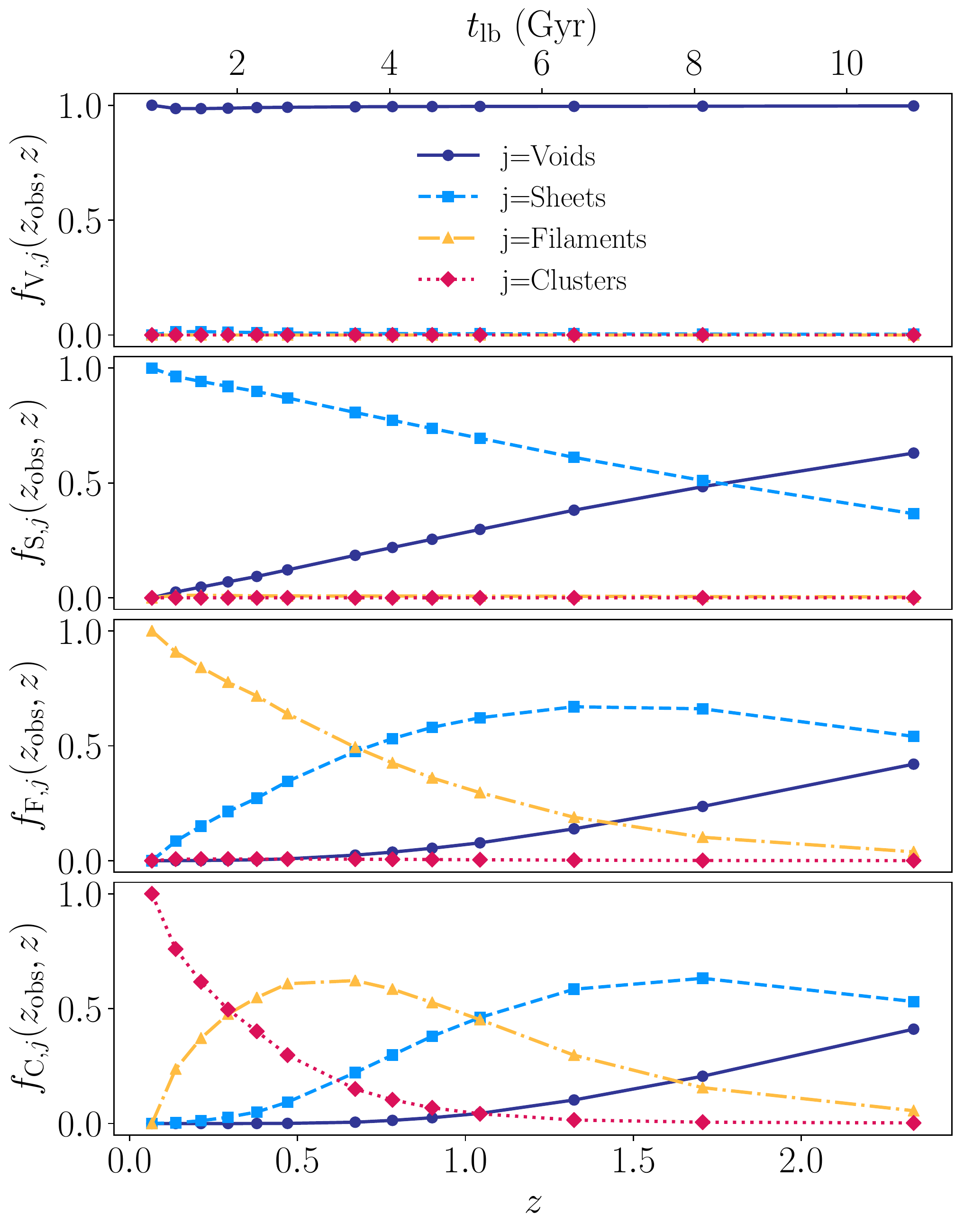}
\caption{Galaxies transport among different types of structures. From top to bottom, the fraction of galaxies observed in voids, sheets, filaments, and clusters, respectively, belonging to different environments $j=\mathrm{V,S,F,C}$ at $z\geq z_\mathrm{obs}$. Top axis marks the corresponding look-back time $t_\mathrm{lb}$.}\label{fig:mass_tr_all}
\end{figure}

\subsection{Environmental history and galaxy properties}\label{sec:Env_gal}

After assessing the accuracy of the cosmic web reconstruction as a function of redshift, we now focus on the past trajectories of individual galaxies through different cosmic structures.
In Section~\ref{sec:accretion}, we show how eFAM allows us to study the variety of environmental histories among galaxies belonging to the same structure type at $z_\mathrm{obs}$. In Section~\ref{sec:gas_fraction}, we exploit this information to investigate whether the observed gas content of cluster members shows the imprint of the time a galaxy spent within a cluster. Finally, in Section~\ref{sec:gas_fraction_fl}, we extend the analysis to filament galaxies.

\vspace{-2mm}
\subsubsection{Galaxy transport among different environments}\label{sec:accretion}

We consider the set of galaxies belonging to different environments at $z_\mathrm{obs}=0.07$ and determine their environmental history, i.e. their host environment as a function of redshift, $z$, by tracing their trajectories back-in-time through the reconstructed cosmic web as detailed in Section~\ref{sec:reconCW}.

Figure~\ref{fig:mass_tr_all} shows the variety of environmental histories among galaxies observed in voids, sheets, filaments, and clusters (top to bottom). 
Different curves represent the fraction of galaxies, $f_{i,j}(z_\mathrm{obs},z)$, detected in the environment of type $i$ at $z_\mathrm{obs}$ and belonging to the structure of type $j$ at $z\geq z_\mathrm{obs}$. 
The colour code is the same as in Figure~\ref{fig:Purity}. Reconstruction suggests that the galaxies observed in voids at $z_\mathrm{obs}=0.07$ have not changed their host environment in the redshift range $0.07<z<2.3$. Conversely, about 40\% of objects detected in sheets have always resided in the same environment, while the remaining 60\% have slowly drifted from voids to sheets in the considered redshift range. Filaments show a more complex formation history. At the highest targeted redshift, i.e., $z=2.3$, future filaments' members are almost equally distributed among sheets and voids. Later, up to $z=1.2$, galaxies originally in voids stream out to sheets at a higher rate than galaxies moving out from sheets to filaments, causing an increase in the fraction of galaxies in sheets between $1.2 < z < 2.3$. Finally, at $z>1.2$, almost all galaxies have left the voids, with the majority of them entering filaments from sheets prior to $z\leq0.7$. 
Clusters members (bottom panel) show the largest variety of environmental histories. Similarly to objects in filaments, at $z=2.3$ future clusters' members are equally distributed among sheets and voids. At later epochs, down to $z=1$, they are gradually poured into filaments after crossing sheets; almost no galaxy have reached a cluster at this time. At $z<1$, galaxies begin to fall into clusters at increasing rate reaching and overtaking the fraction of objects in filaments by $z\leq 0.3$. 

Overall, eFAM traces the hierarchical paradigm of structure formation \citep{zel'dovic1970}, predicting the flow of galaxies from low to high-density regions. The bottom-up formation scenario of clusters and filaments has been already assessed using the same statistics, i.e. $f_{i,j}(z_\mathrm{obs},z)$, in \cite{Cautun2014MNRAS.441.2923C} by applying the Nexus+ \citep{Nexus+2013} cosmic web classifier on a series of simulated snapshots extracted from an $N$-body cosmological simulations. In contrast, Nexus+ suggests the up-to-bottom formations of voids and sheets tracing the 20 per cent of the mass belonging to voids and sheets at $z=0$ back to sheets and filaments, respectively, at higher redshifts. We attribute this discrepancy to the different choice of the threshold $\mu_\mathrm{th}$ used in the web classification, which plays a crucial role in discriminating between non-collapsing structures.

\vspace{-2mm}
\subsubsection{Gas fraction and environmental history in cluster galaxies}\label{sec:gas_fraction}

\begin{figure}
\centering
\includegraphics[width=1.\columnwidth]{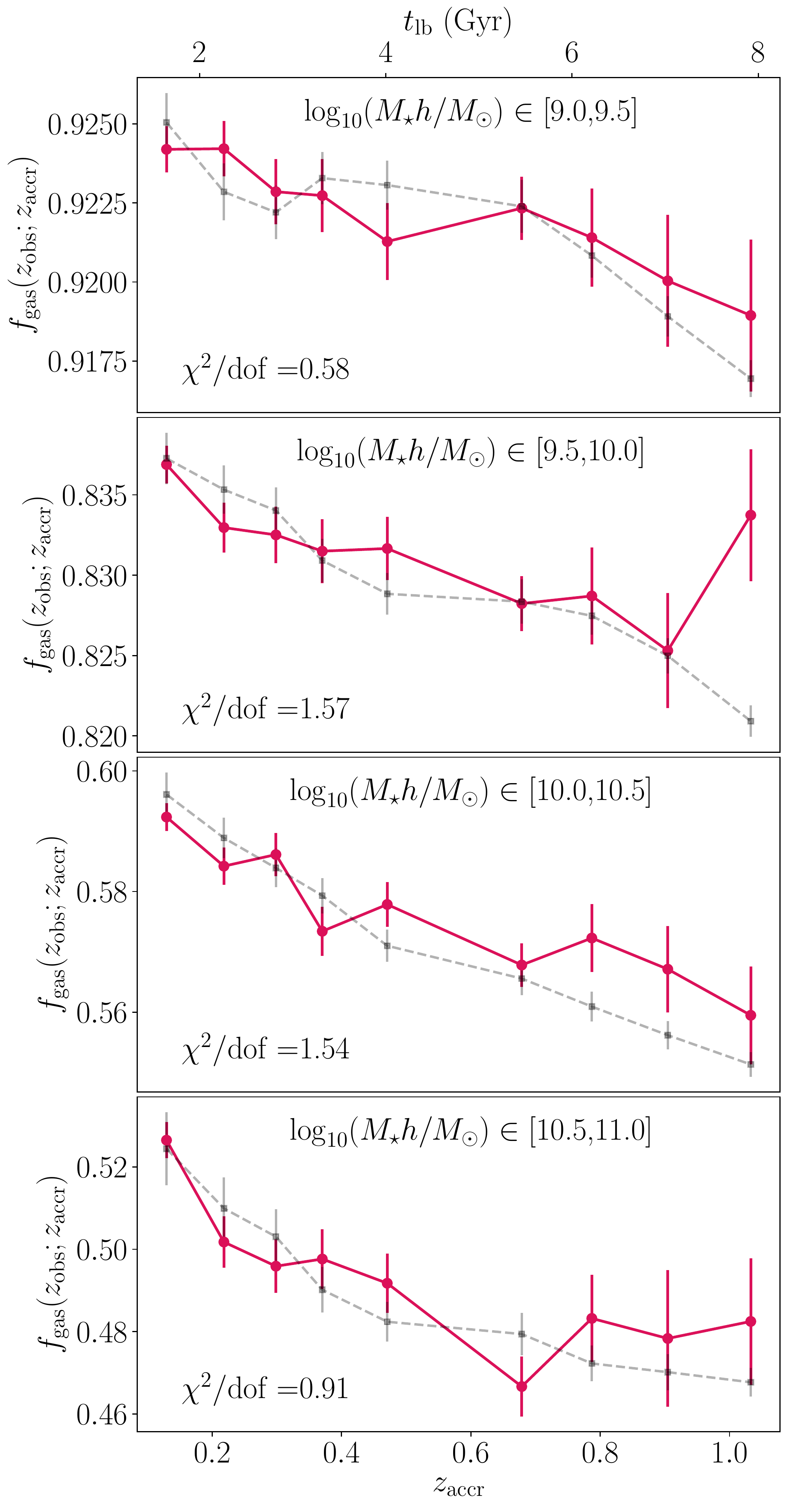}
\caption{Gas fraction at $z_\mathrm{obs}=0.07$ of cluster galaxies as a function of their redshift of accretion. On the y-axis, $f_\mathrm{gas}(z_\mathrm{obs};z_\mathrm{accr})$ represents the median gas fraction estimated at $z_\mathrm{obs}$ averaged over all cluster members accreting at $z=z_\mathrm{accr}$. Red-solid lines mark the results of eFAM reconstruction, while grey dashed lines are predictions based on \magnet simulation. Error bars are the standard deviation of the measurements. From top to bottom, the relation is shown for increasing bins of stellar mass. The agreement between the reconstructed and simulated trends is evaluated by means of the $\chi^2$ distance.}\label{fig:CL_gasfrac}
\end{figure}

It is well established that galaxies belonging to dense environments such as clusters or groups of galaxies have on average a lower gas content than their counterparts in the field \citep[e.g. see][for a reiew]{Boselli2021arXiv210913614B}. However, it is yet unclear whether for observed galaxies we can detect any dependence of the measured gas content of cluster members on the time they have spent in a cluster, or equivalently on their redshift of accretion to the cluster, $z_\mathrm{accr}$. In this Section, we will show that via eFAM we can successfully address this question by reconstructing the environmental histories of cluster galaxies.
We proceed as follows: for each uncompressed object (see Section~\ref{sec:methods:FoG} for details on compression) detected in a cluster at $z_\mathrm{obs}$, \emph{i)} we estimate $z_\mathrm{accr}$ as the lowest redshift at which its reconstructed trajectory intersects a filament, a sheet, or a void, and \emph{ii)} we compute its gas fraction as $f_\mathrm{gas}(z_\mathrm{obs})=M_\mathrm{gas}/(M_{\star}
+M_\mathrm{gas})$, where $M_{\star}$ and $M_\mathrm{gas}$ are the galaxy stellar mass and total gas content (hot and cold), respectively and \emph{iii)} we evaluate the dependence of the measured gas fraction on $z_\mathrm{accr}$, $f_\mathrm{gas}(z_\mathrm{obs};z_\mathrm{accr})$, by computing the median of $f_\mathrm{gas}(z_\mathrm{obs})$ in bins of $z_\mathrm{accr}$ corresponding to the reconstructed catalogues listed in Section~\ref{sec:data}. In Figure~\ref{fig:CL_gasfrac} we plot the reconstructed $f_\mathrm{gas}(z_\mathrm{obs};z_\mathrm{accr})$ with their uncertainties, the latter computed as the standard deviation on the median values (orange-solid lines and error bars). To ensure the statistical significance of our estimates, we limit the analysis to $z_\mathrm{accr}\leq 1.2$, given that only a small fraction ($< 5$ per cent) of galaxies have reached cluster structures at higher redshifts (see Figure~\ref{fig:mass_tr_all} bottom panel).
Here, to account for any dependence of $f_\mathrm{gas}$ on stellar mass, we further divide the galaxy sample in four bins of stellar mass, namely $\log_{10}(M_\star h/M_\odot)\in [9,9.5]$, $[9.5,10]$, $[10,10.5]$, and $[10.5,11]$ (top to bottom). 
Independently of the mass bin considered, $f_\mathrm{gas}(z_\mathrm{obs};z_\mathrm{accr})$ shows a clear decreasing trend. This trend, in agreement with the prediction based on \magnet simulations (grey lines in Figure~\ref{fig:CL_gasfrac}, see Section~\ref{sec:validation} for the details on validation), highlights the role of the environment in driving galaxy evolution.
As a consequence of the interaction with the high density environment, galaxies residing within a cluster keep losing their gas content, so that galaxies that have spent more time in the cluster (higher $z_\mathrm{accr}$) tend to have on average less gas per unit of baryonic mass. The higher fraction of gas loss is measured within the first few Gyrs after accretion, later on the gas loss is less efficient and galaxies that have accreted at higher redshifts show similar $f_\mathrm{gas}$. Moreover, the lower the mass bin considered the shorter is the time scale over which such a phenomenon occurs, as we expect.
Finally, Figure~\ref{fig:CL_gasfrac} suggests that the efficiency of environmental processing increases with increasing stellar mass, in apparent contrast with the expectation that low-mass galaxies are more affected by the environment than more massive objects. However, we emphasise that as opposed to previous studies based on observations \citep[e.g.][]{Cortese2011}, we are analysing \emph{i)} the total (cold plus hot) gas fraction of the galaxies, rather than solely the cold component, and \emph{ii)} we are labelling as cluster galaxies all the objects within $2-3$ times the virial radius and not belonging to the virialised core of the structure (see Section~\ref{sec:methods:FoG} for details on the compression of cluster cores), thus we do not sample the cluster regions where environmental effects are more efficient. Given these considerations, we establish the reliability of our method by direct comparison with \magnet simulation.

\begin{figure}
\centering
\includegraphics[width=1.\columnwidth]{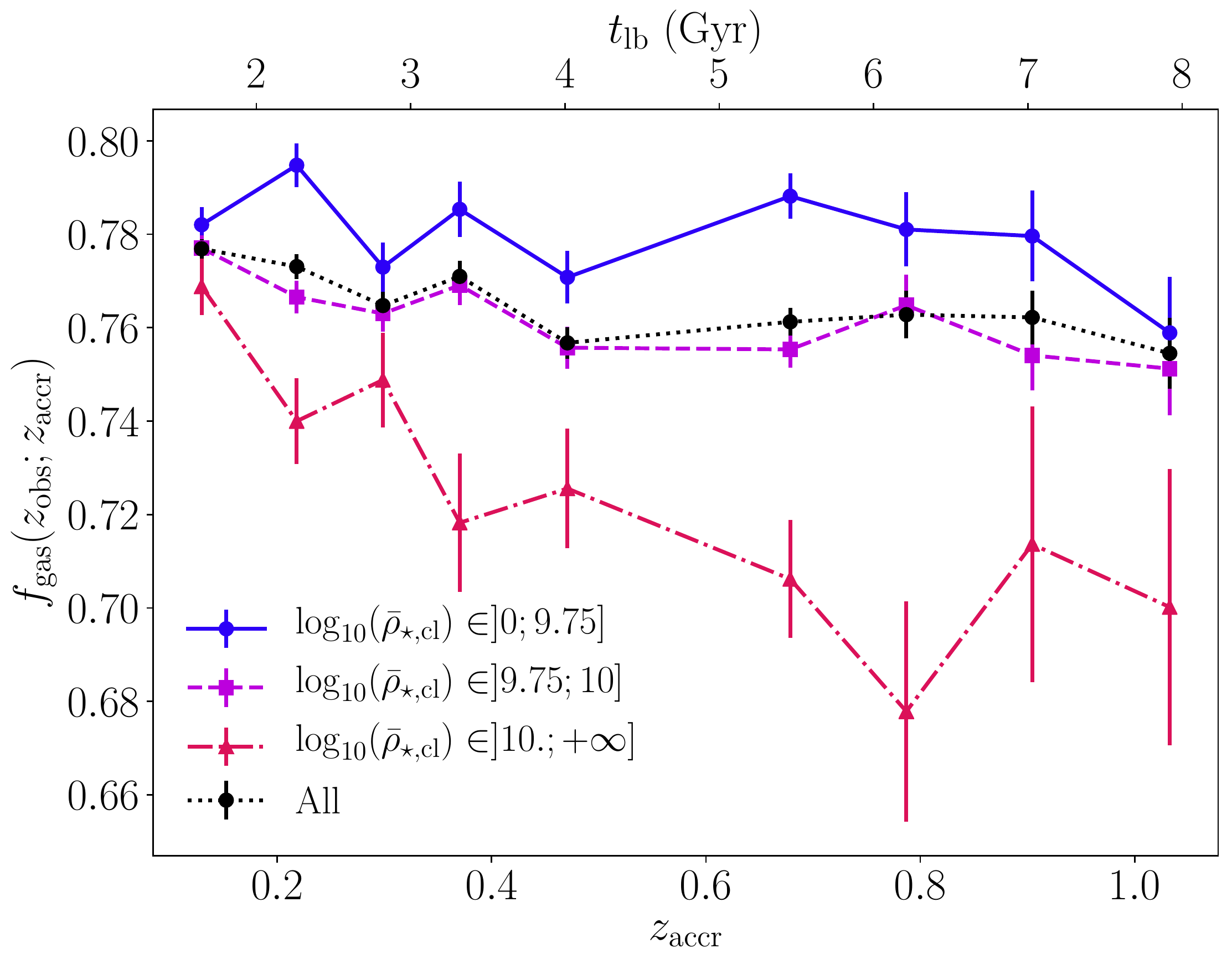}
\caption{Gas fraction at $z_\mathrm{obs}=0.07$ of cluster galaxies as a function of their redshift of accretion for different stellar mass densities of clusters.
$f_\mathrm{gas}(z_\mathrm{obs};z_\mathrm{inf})$ is computed as in Figure~\ref{fig:CL_gasfrac}. The relation is shown for different values of median stellar mass density of the host cluster, $\bar{\rho}_{\star,\mathrm{cl}}$ in units of $(h^2 M_\odot \mathrm{Mpc}^{-3})$.}\label{fig:gas_rich}
\end{figure}

As next step, we explore whether our findings are sensitive to the mean stellar mass density of the cluster, $\bar{\rho}_{\star,\mathrm{cl}}$.
To estimate $\bar{\rho}_{\star,\mathrm{cl}}$, we group all contiguous cells of type cluster into a single structure of volume $V_\mathrm{cl}$ and take the ratio between the sum of the stellar masses of their members and $V_\mathrm{cl}$. As in the previous analysis, we compute $f_\mathrm{gas}(z_\mathrm{obs};z_\mathrm{accr})$ in bins of $\bar{\rho}_{\star,\mathrm{cl}}$.
Figure~\ref{fig:gas_rich} illustrates the median gas fraction as a function of $z_\mathrm{accr}$ for galaxies belonging to clusters in the first (blue-solid line), second (purple-dashed line), and third (red dot-dashed line) 33rd percentile of stellar mass density. As a reference reference, we also show in black-dotted line the median gas fraction for all galaxies. For this analysis, we can not provide a direct comparison with predictions based on \magnet simulation since our fiducial model illustrated in Section~\ref{sec:validation} is based on the assumption that galaxies preserve their bin in $M_\star$ throughout the considered redshift range. This assumption does not hold if instead we bin in cluster density, which steeply increases approaching $z_\mathrm{obs}$. 

The results provided by eFAM agree with the dependence of the observed gas fraction on the local density of clusters seen in nearby structures \citep{2014BoselliA&A...564A..66B}, confirming that the denser the environment the more efficient the gas loss. 

\vspace{-2mm}
\subsubsection{Gas fraction and environmental history in filaments}\label{sec:gas_fraction_fl}
\begin{figure}
\centering
\includegraphics[width=1.\columnwidth]{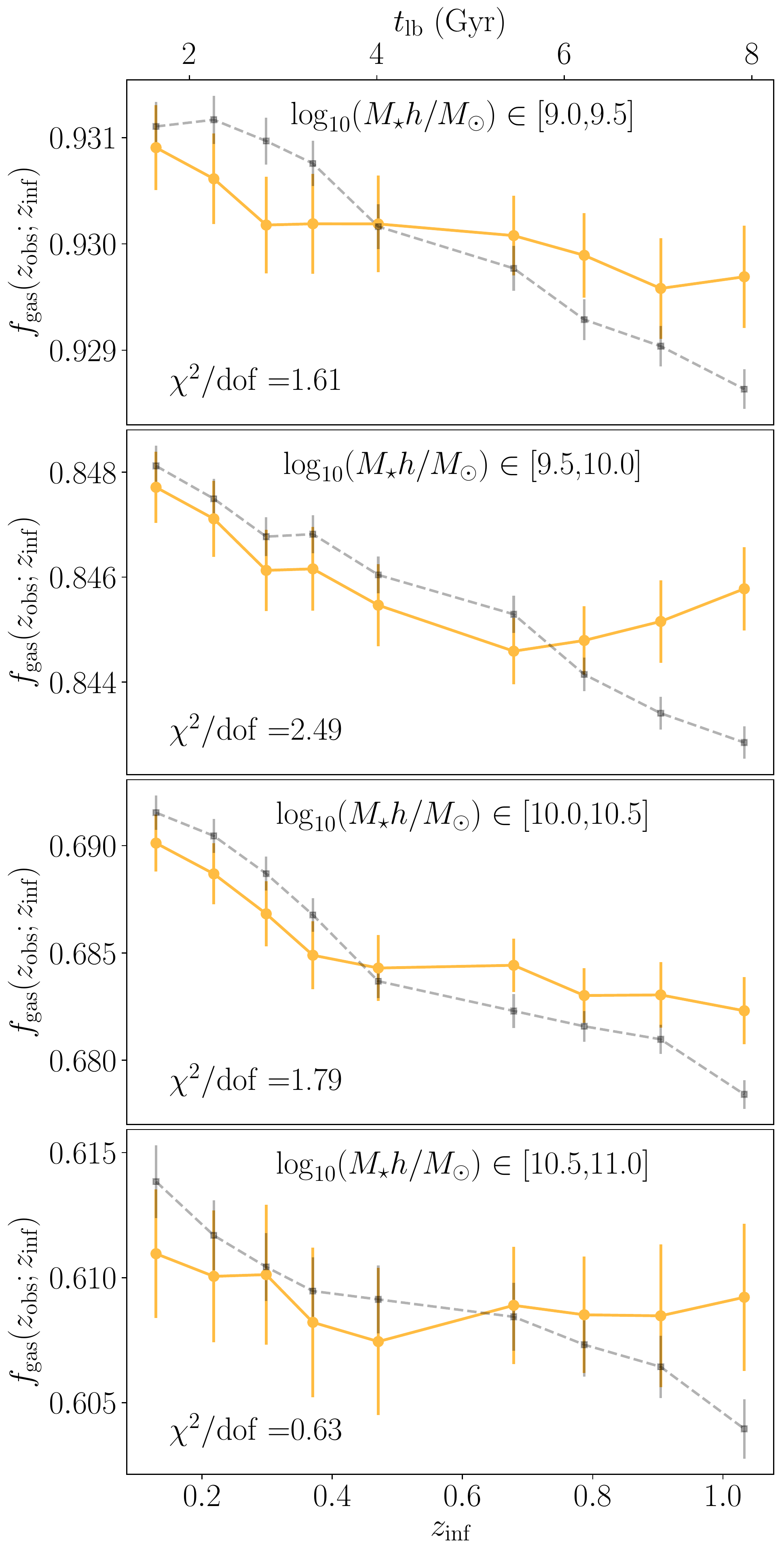}
\caption{As in Fig. \ref{fig:CL_gasfrac}, however this time the relation is shown for galaxies in filaments as a function of their redshift of infall.}\label{fig:FL_gasfrac}
\end{figure}

We conclude our study by investigating whether the observed gas content of filament galaxies shows an environmental dependence similar to that of the cluster members. 
To this end, we consider the set of galaxies detected in a filament at $z_\mathrm{obs}$, determine their redshift of infall into the filament, $z_\mathrm{inf}$, as the lowest redshift at which they are tracked back into a sheet or a void, and construct $f_\mathrm{gas}(z_\mathrm{obs};z_\mathrm{inf})$ as in Section~\ref{sec:gas_fraction}. When analysing filament galaxies, we identify two distinct populations: satellite galaxies belonging to non-compressed small groups, i.e with less than five members (Section~\ref{sec:methods:FoG}), and field galaxies. These two families are expected to experience different environmental influences, with galaxies in groups being processed more efficiently than their counterparts in the field \citep[e.g.][]{Cortese2006,DeLucia2012,Castignani2021}. To highlight the effect of filaments, beyond that of groups, we focus on field galaxies alone.

Figure~\ref{fig:FL_gasfrac} (orange solid lines) shows the reconstructed $f_\mathrm{gas}(z_\mathrm{obs};z_\mathrm{inf})$ and its variance evaluated in four stellar mass bins of galaxies. Up to $M_{\star}=10^{10.5}h^{-1}M_{\odot}$, $f_\mathrm{gas}(z_\mathrm{obs};z_\mathrm{inf})$ decreases as a function of $z_\mathrm{inf}$, in fair agreement ($\chi^2/dof$ distance ranging from 0.63 to 2.49) with \magnet-based predictions (dashed grey lines). Here, the gas depletion in simulated galaxies is obtained by adapting Equation~\ref{eq:model} to take into account the transition between filaments and sheets, and assuming that most of filament galaxies have been poured from sheets in the interval $0.07\leq z \leq 1.2$, as suggested by Figure~\ref{fig:mass_tr_all}. The recovered trend reveals that galaxy processing occurs in filaments as in clusters and that galaxies are gradually gas depleted as they flow through filaments. However, the maximum variation of $f_\mathrm{gas}(z_\mathrm{obs};z_\mathrm{inf})$ observed for galaxies in filaments is about 25 per cent of that estimated among cluster members, showing that environmental effects are less efficient in lower density regions, as expected. As with cluster galaxies, we find that the reconstructed $f_\mathrm{gas}(z_\mathrm{obs};z_\mathrm{inf})$ in the mass bin $\log_{10}(M_{\star} h/M_\odot)\in[10,10.5]$ tends to increase at high redshift. This overestimation may arise either from an ambiguous classification of the cosmic web at high redshift, as suggested by the decreasing trend of structures purity (see Figure~\ref{fig:Purity}), or from the simulated galaxy evolution itself which shows an excess of galaxies in the considered mass bin. We will further investigate this issue in future works. 
In contrast to cluster galaxies (Figure~\ref{fig:CL_gasfrac}), the most massive objects in filaments are the least responsive to environmental effects and show only 5 per cent of the variation observed in their counterparts in clusters. 
We can speculate that at higher redshifts, the temperature and densities in filaments are not high enough to perturb significantly galaxies with stellar masses $10^{10.5}h^{-1}M_\odot$.

\vspace{-2mm}
\subsubsection{Validation against simulations}\label{sec:validation}
\begin{figure}
\centering
\includegraphics[width=1.\columnwidth]{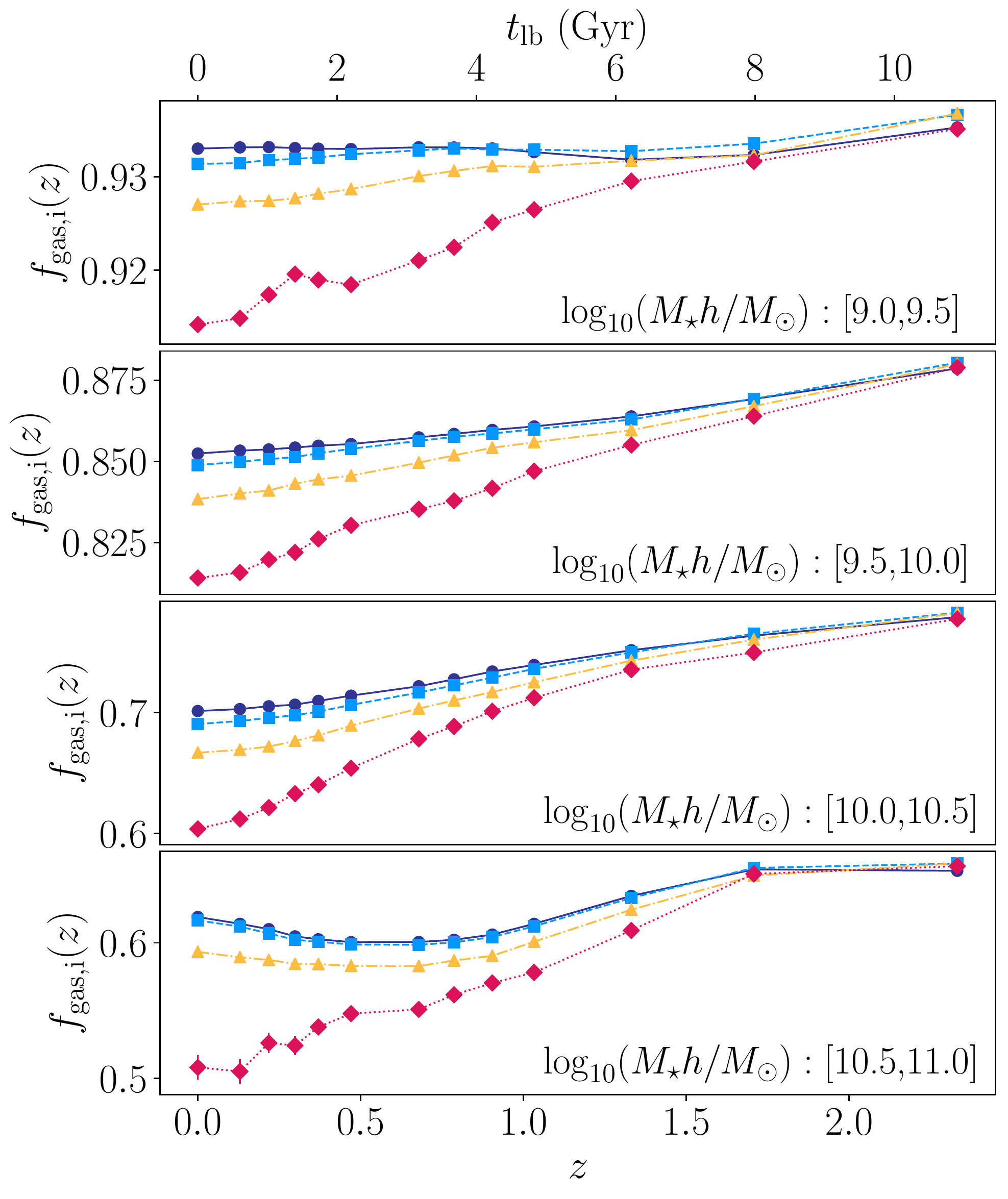}
\caption{Evolution of gas fraction as a function of redshift within different environments. Different panels correspond to different galaxies stellar mass bins. Colour code as in Figure~\ref{fig:mass_tr_all}}\label{fig:sim_fgas}
\end{figure}

In the absence of the galaxies merger trees, we validate the results presented in the previous section by constructing a model for $f_\mathrm{gas}(z_\mathrm{obs};z_\mathrm{accr})$ using the information available in the simulation.
We consider the simulated snapshots listed in Section~\ref{sec:data}. 
We estimate the evolution of $f_\mathrm{gas}(z)$ within different structures types averaging over the gas fraction of galaxies belonging to voids, sheets, filaments, and clusters, respectively as identified by the T-web classifier. 

The results are illustrated in Figure~\ref{fig:sim_fgas} using the same colour code as in Figure~\ref{fig:mass_tr_all}. As in Figure~\ref{fig:CL_gasfrac}, we divide objects according to their stellar mass. Regardless of the stellar mass bin, $f_\mathrm{gas}$ evolves differently in different environments; as the redshift decreases, the difference between the $f_\mathrm{gas,i}$ with $i=\{\mathrm{V},\mathrm{S},\mathrm{F},\mathrm{C}\}$ increases, with galaxies in clusters showing systematically lower gas fractions than their counterparts in filaments. Concurrently, the gas loss rate, $\partial f_\mathrm{gas}/\partial z$, in filaments is higher than in sheets and voids. $f_\mathrm{gas}$ progressively decreases with increasing $M_\star$ (top to bottom).

These results suggest that at relatively high redshifts, $z_\mathrm{in}>1.7$, galaxies with comparable $M_\star$ have similar $f_\mathrm{gas}$. Then, as they evolve within the cosmic web, they loose gas at a rate, $\partial f_\mathrm{gas,i}/\partial z$, that depends solely on their stellar mass and hosting environment $i$ at the analysed redshifts.
We construct a model of $f_\mathrm{gas}(z_\mathrm{obs};z_\mathrm{accr})$ based on these considerations. Assuming that most of galaxies are accreted from filaments as hinted by the bottom panel of Figure~\ref{fig:mass_tr_all}, and that a galaxy preserves its mass bin throughout the considered redshift range, we define $f_\mathrm{gas}(z_\mathrm{accr};z_\mathrm{obs})$ as
\begin{equation}\label{eq:model}
    f_\mathrm{gas}(z_\mathrm{obs};z_\mathrm{accr})
    = f_\mathrm{gas}(z_\mathrm{in}) + \int_{z_\mathrm{in}}^{z_\mathrm{accr}}dz \frac{\partial f_\mathrm{gas,F}}{\partial z}+ \int_{z_\mathrm{accr}}^{z_\mathrm{obs}}dz\frac{\partial f_\mathrm{gas,C}}{\partial z},
\end{equation}
where $f_\mathrm{gas}(z_\mathrm{in})$ is the initial galaxy gas fraction computed as the mean of $f_\mathrm{gas}$ at $z_\mathrm{in}=2.33$ when no distinction between structures is considered, and the two integrals represent the total amount of gas lost during the time spent by a galaxy within a filament first and inside a cluster afterwards.
To model the time-integrated gas depletion in galaxies belonging to a filament at $z_\mathrm{obs}$, we modify Equation~\ref{eq:model} to account for the transition between sheets and filaments rather then the accretion from filaments to clusters.
The predicted $f_\mathrm{gas}(z_\mathrm{obs};z_\mathrm{accr})$ is shown by grey lines in Figures~\ref{fig:CL_gasfrac} and \ref{fig:FL_gasfrac} for clusters and filaments, respectively. 

We assess the reliability of eFAM prediction by computing the $\chi^2/dof$ distance between simulated and reconstructed trends in each mass bin. Since we are considering the galaxy gas fraction measured at the observed redshift, the effect of a faulty reconstruction (e.g. an inaccurate estimation of galaxies' peculiar velocities) would be a wrong estimation of $z_\mathrm{accr}$, thus misplacing the points along the horizontal axis in the $(z_\mathrm{accr},f(z_\mathrm{obs},z_\mathrm{accr}))$ plane and erasing the underlying trend. This is not the case. In cluster galaxies, we find $\chi ^2/dof<1$ for the mass bins $\log_{10}(M_{\star} h/M_\odot)\in[9,9.5],[10.5,11]$ and $\chi ^2/dof<1.6$ for $\log_{10}(M_{\star} h/M_\odot)\in[9.5,10],[10,10.5]$ denoting a fair agreement between eFAM prediction and the simulated trend. The predictions for filament galaxies are slightly less accurate with $\chi ^2/dof<1.8$ for $\log_{10}(M_{\star} h/M_\odot)\in [9,9.5],[9.5,10],[10.5,11]$ and $\chi ^2/dof=2.5$ in $\log_{10}(M_{\star} h/M_\odot)\in [10,10.5]$. As discussed in Section~\ref{sec:discussion}, we plan to further investigate the mass bin $\log_{10}(M_{\star} h/M_\odot)\in [10,10.5]$ in future work by applying eFAM to a different hydrodinamical simulation. These results suggest that the current implementation of eFAM is accurate enough to fairly recover the environmental dependence of the observed properties of both cluster and filament galaxies. 


\vspace{-2mm}
\section{Discussion}\label{sec:discussion}
In the previous sections, we have shown that eFAM reconstruction is a powerful tool to model the redshift evolution of the cosmic web and extract the environmental history of the observed mass tracers, providing the opportunity for a deterministic study of the environmental dependence of observed galaxy properties.
Here, we prepare the ground for future applications by establishing the reliability domain of the method also discussing the possibility of applying eFAM to the study of galaxy pre-processing.

The efficiency and reliability of back-in-time reconstruction techniques depend, in general, on the number density of the sample on which they are applied. On the one hand, very sparse samples lack the minimum information to recover the non-linear evolution of structures. On the other hand, the high degree of non-linearity required to model high-density regions is difficult to achieve. In this work, we test the performances of eFAM on a high-density catalogue mimicking the observed fields of low-redshift spectroscopic galaxy surveys such as GAMA \citep{GAMA2011MNRAS.413..971D} and VIPERS \citep{VIPERS2014A&A...566A.108G}, and show that, after the compression of the core of the clusters, eFAM recovers the correct redshift evolution of the cosmic web when applied to a sample with number density up to $\bar{n}= 10^{-2}h^3\mathrm{Mpc}^{-3}$. The accuracy of eFAM on lower density catalogues has been instead investigated in previous works. In \cite{eFAM22021MNRAS.tmp..398S}, we applied eFAM to the lowest redshift bin, $0.3<z<0.5$, of SDSS-DR12 COMBINED \citep{sdss_dr12} galaxy sample with $\bar{n}=3\times10^{-4}h^3\mathrm{Mpc}^{-3}$ successfully retrieving both its velocity and past-density field. In \cite{eFAM12019MNRAS.484.3818S}, we tested the eFAM algorithm on an extremely sparse dark matter halo catalogue with $\bar{n}\sim10^{-5}h^3\mathrm{Mpc}^{-3}$ showing that eFAM is able to recover the linear real-space clustering statistics even at densities as low as the ones of eBOSS quasar samples \citep{eBOSS}.
These results demonstrate the robustness of eFAM accuracy with respect to the number density of the sample and support its application to future, lower-density and/or higher-redshift spectroscopic surveys such as Euclid \citep{Euclid2020}, designed to reach number densities of $4.8\times 10^{-4} h^3\mathrm{Mpc}^{-3}$ in the redshift interval $0.9<z<1.9$ (wide field), The Prime Focus Spectrograph \citep{PFS2014}, expected to reach a mean number density of about $4.7 \times 10^{-4} h^3\mathrm{Mpc}^{-3}$ between $0.6 < z <2.4$, and the MauneaKea Spectroscopic Explorer \citep{MSE2019}, intended to attain $\bar{n} = 1.3\times 10^{-4} h^3\mathrm{Mpc}^{-3}$ in the redshift range $1.6< z <4$.   


One of the open questions in galaxy evolution is understanding galaxy pre-processing, defined here as the environmental effect cluster members feel within filaments and groups before their accretion into clusters. 
In Section~\ref{sec:Env_gal}, we have shown that eFAM captures the effect of the environment on the observed properties of both cluster and filament galaxies, hinting at the possibility of combining the two analyses to investigate pre-processing. 
However, this kind of analysis demands the estimation of the cosmic web on sub-Mpc scales in order to properly resolve filaments in the interior of clusters; a resolution that cannot be reached with T-web implementation adopted in this work which requires to interpolate the density field on a relatively wide grid to limit the effect of Poisson noise. In this work, we set the step size of the grid covering the \magnet simulated volume to $1.68h^{-1}\mathrm{Mpc}$, corresponding to half the mean inter-particle distance at $z=0.07$.
We plan to study the galaxy pre-processing in future work by pairing eFAM to a finer web classifier.

Finally, as discussed in Section~\ref{sec:gas_fraction}, the direct comparison with observations of galaxy evolution will be made possible by accessing more detailed information on the observed galaxy properties such as the different gas phases. In particular, thanks to the wide coverage and angular resolution of upcoming surveys as the Square Kilometer Array \citep{SKA}, the analysis of the cold atomic hydrogen phase, HI, will play a crucial role in the understanding of galaxy processing in clusters.
In future works, we plan to combine the detailed information on observed galaxies with an improved model of the non-linear dynamics in high-density regions to study the environment-induced gas depletion within different structures.

\vspace{-2mm}

\vspace{-2mm}
\section{Summary and Conclusions}\label{sec:summary}

We presented the first application of the extended Fast Action Minimization method (eFAM) to the study of time-integrated environmental effects on observed galaxies properties. 

We based our benchmark analysis on the cosmological hydrodynamical simulation Magneticum Pathfinder and treated the sample of simulated galaxies at $z_\mathrm{obs}=0.07$, and with $M_\star\geq 10^9 h^{-1}M_\odot$, as the observed catalogue. 
By applying eFAM to the observed galaxy distribution in redshift-space, we recovered the past trajectories of the observed tracers and used this information to build a series of three-dimensional density fields at high redshifts. We then applied the T-web classifier to each reconstructed field to estimate the redshift evolution of the cosmic web and infer the environmental history of individual galaxies by combining their trajectories with the large-scale information.
We validated our results by performing a similar analysis on the higher redshift snapshots of the simulations.

The main results of our work can be summarised as follows:
\begin{itemize}
    \item \emph{Cosmic web evolution} - eFAM accurately recovers the evolution of the cosmic web through time. At the statistical level, this result is confirmed by the $1\sigma$ agreement between the volume filling fraction of different structures types as measured in the reconstructed and simulated galaxy catalogues. We assessed the local accuracy of the reconstructed web by measuring the purity, $P$, of structures at the cells of a cubic grid enfolding the sample. Up to $z=1.2$, clusters have $0.58<P<0.93$, filaments range in $0.90<P<0.99$, sheets have $0.78<P<0.92$, and voids are best identified with $0.90<P<0.92$. As the redshift increases, the differences in number density and bias between the reconstructed tracers and the simulated galaxies increase, making it more difficult to compare the two fields and the purity of clusters decreases to 60 per cent. This missclassification does not affect the quality of the reconstructed environmental histories of cluster members since less than five percent of those galaxies belong to a cluster at $z>1.2$. 
    
  \item \emph{Environmental effect on galaxy evolution} - when coupled with a cosmic web classifier, eFAM allows us to retrieve the variety of environmental histories of galaxies observed in similar environments and corroborates the hierarchical paradigm of structure formation. Additionally, it offers the unique opportunity to investigate whether the observed properties of galaxies are affected by their environmental history. Focusing on the observed gas fraction, $f_\mathrm{gas}$, of cluster members, we found that the average gas per unit of baryonic mass decreases as a function of the redshift of action to the cluster. This trend, in agreement with prediction based on \magnet simulation, locates the higher fraction of gas loss within the first few Gyrs after accretion and show that galaxies that have accreted at higher redshifts have similar $f_\mathrm{gas}$. We then extended the analysis to galaxies belonging to filaments at the observed redshift and found that galaxies are subject to pre-processing, suffering gas depletion in filaments as they do in clusters, although the effect is reduced. Finally, in contrast with cluster galaxies, the most massive objects, i.e. with $\log_{10}(M_{\star} h/M_\odot)\in[10,10.5]$, in filaments are the least sensitive to environmental interactions suggesting the mean density in filaments to be insufficient for them to experience the influence of the environment they reside in.
\end{itemize}

The results obtained in this work prove eFAM to be a useful tool for the deterministic study of the environmental dependence of observed galaxy properties, offering a complementary approach to the one based on light-cone observations.




\vspace{-2mm}
\section*{Acknowledgements}
The authors acknowledge the anonymous referee for the very useful comments that improved the quality of the paper. ES acknowledges Julian Bautista for the careful review of the second version of the draft and Clotilde Laigle, Henry Mc Cracken, and Marko Shuntov for the very insightful discussions during the early preparation of the paper. The authors acknowledge Alessandro Boselli and Enzo Branchini for a critical review of the results presented in this work. The authors acknowledge Klaus Dolag for clarification and support on Magneticum simulation, and Agnieska Pollo, Kasia Ma\l ek, and Olga Cucciati for useful comments. AL is supported by Fondazione Cariplo, grant No 2018-2329. CS is partially supported by the Programme National Cosmology et Galaxies (PNCG) of CNRS/INSU with INP and IN2P3, co-funded by CEA and CNES. KK acknowledges support from the DEEPDIP project (ANR-19-CE31-0023). The project leading to this publication has received funding from Excellence Initiative of Aix-Marseille University - A*MIDEX, a French "Investissements d'Avenir" programme (AMX-19-IET-008 - IPhU). In Memoriam of Amalia Pisanelli.

\vspace{-2mm}
\section*{Data Availability}
The \magnet simulations used in this work are publicly available at http://www.magneticum.org. The T-web catalogues and the eFAM reconstructed snapshots are available from the authors, upon request.


\bibliographystyle{mnras}
\bibliography{CWrec} 



\appendix
\section{Accuracy of eFAM reconstruction}\label{app:velocities}

\begin{figure}
\centering
\includegraphics[width=0.9\columnwidth]{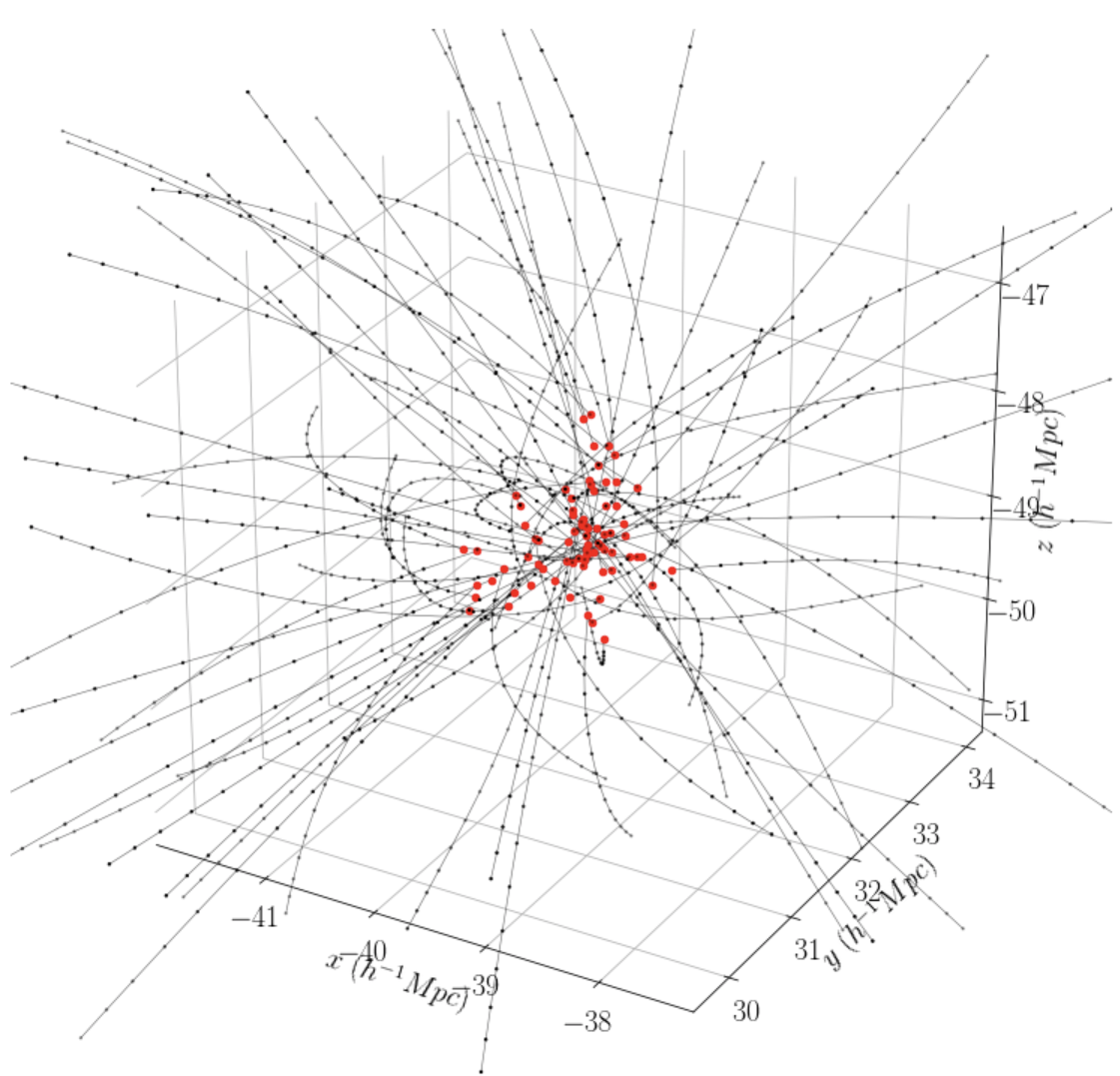}
\caption{eFAM reconstructed orbits of cluster members. Red dots are the galaxies positions at the observed redshift while black points mark the back-in-time positions along the orbit at 14 redshifts in the interval $0.07\leq z\leq 1.1$. As expected, some of the reconstructed orbits strongly deviate from the linear path predicted by the Zel’dovich solution.
}\label{fig:NLtraj}
\end{figure}

In this Appendix we explore the reliability of eFAM reconstruction within different environments and compare the results with the Zel'dovich reconstruction \citet[][hereafter quoted as ZA]{ZArec}. 

To best approach the global minimum of the action, eFAM uses the Zel’dovich approximation as the first guess in the action minimisation. This approach does not restrain the eFAM solution to the linear regime and serves only to guide the minimization. Figure~\ref{fig:NLtraj} offers a striking example of the non-linearity of eFAM trajectories within dense regions, highlighting the difference between eFAM and ZA solutions. Here, black lines are reconstructed orbits of the members of a dense cluster in the \magnet galaxy distribution, red dots mark the galaxy position at the observed redshift, and black points are the back-in-time positions in the redshift range $0.07<z< 1.1$.

\begin{figure*}
\centering
\includegraphics[width=1.\textwidth]{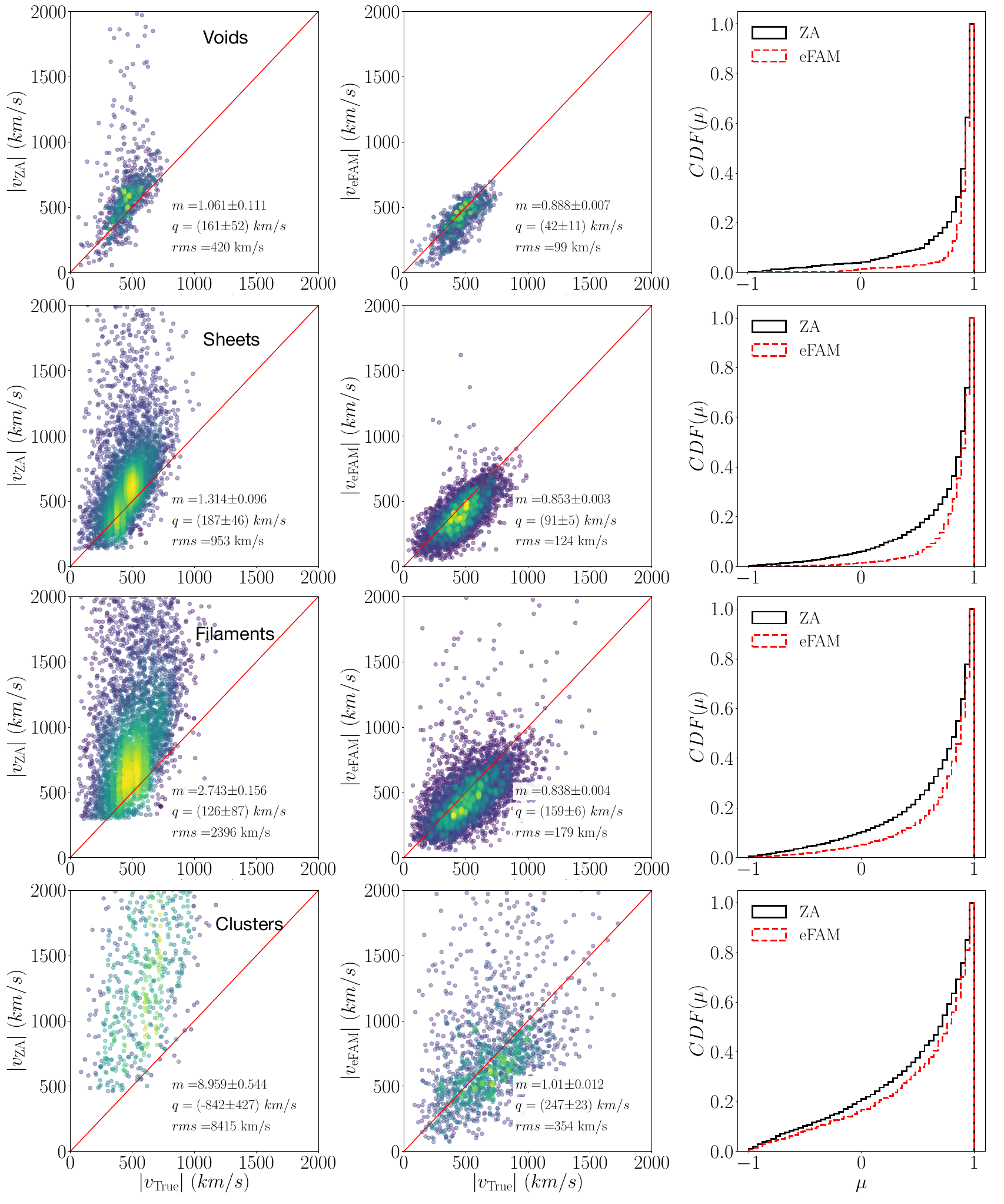}
\caption{Accuracy of reconstruction in different environments: comparison between eFAM and Zel'dovich (ZA) solution. The left and mid columns are the point-to-point comparison between the ``true'' $N$-body and reconstructed amplitudes of peculiar velocities as obtained via ZA and eFAM, respectively. $m$ and $q$ are the results of the linear regression $|\boldsymbol{v}_\mathrm{eFAM}|=m|\boldsymbol{v}_\mathrm{eFAM}|+q$, $rms$ is the dispersion with respect to the best fit values, and the red diagonals mark the perfect reconstruction, $m=1$ and $q=0$. The right column shows the cumulative distribution function of the cosine of the angle between the ``true'' and simulated velocity vectors, $CDF(\mu)$. A perfect reconstruction would yield $\mu=1$. Different rows refer to different environments, from top to bottom: voids, sheets, filaments, and clusters.
}
\label{fig:eFAM_vs_ZA}
\end{figure*}
In Figure~\ref{fig:eFAM_vs_ZA}, we compare the eFAM and ZA solutions within different environments (voids, sheets, filaments, and clusters,  top-to-bottom lines) by adopting the standard metric for the evaluation of reconstruction performances, i.e. the velocity-velocity comparison \citep[see e.g.][]{NusserBranchini2000,Branchini2002,eFAM12019MNRAS.484.3818S}. 
On the left and middle columns, we show the point-to-point comparison between the ``true''  $N$-body and reconstructed amplitudes of peculiar velocities as modelled by ZA and eFAM, respectively. For each comparison, we report the results of the linear regression $|\boldsymbol{v}_\mathrm{eFAM}|=m|\boldsymbol{v}_\mathrm{eFAM}|+q$ and the rms-variance of velocity differences, $rms$. A perfect reconstruction corresponding to slope $m=1$, offset $q=0$, and vanishing $rms$ (red line). The right column illustrates the cumulative distribution function (CDF) of the cosine, $\mu$, of the angle between the and the reconstructed velocity vectors. A perfect reconstruction corresponds to $\mu=1$. 
ZA performs best within voids, fairly modelling the mean amplitude and direction of the velocity vectors. Still, when moving towards denser regions, ZA progressively overestimates $|\boldsymbol{v}|$ and lowers the accuracy of the inferred orientations. In voids, sheets, and filaments, the eFAM solution is superior to ZA in recovering both the amplitude and the direction of the simulated velocity vectors. 
Within clusters, eFAM struggles to retrieve the correct velocity orientation, still being able to capture its mean amplitude.
This is an expected result. Within virialised or highly non-linear structures, galaxies lose information about their past orbits and follow a random, thermal motion with a velocity dispersion set by the mass of the host cluster. Differently from the amplitude, the velocity orientation is, therefore, not uniquely constrained.

For sake of comparison between eFAM and other Least Action Principle (LAP) methods as for their reliability of recovering $N$-body or observed velocities or initial density field, see e.g. \cite{Branchini1994,Dunn1995, Goldberg2000, Sharpe2001,Romano2005A&A...440..425R, NAM, Shaya2017}. Similar tests, including the comparison with ZA, have been performed in studies not based on LAP, e.g. \cite{Kitaura2008,Kitaura2013, Leclercq2015JCAP...06..015L,Lavaux2016,Wang2022}.

\section{Simulated vs. reconstructed two-point clustering }\label{app:xi_comparison}

\begin{figure*}
    \centering
    \includegraphics[width=.9
    \textwidth]{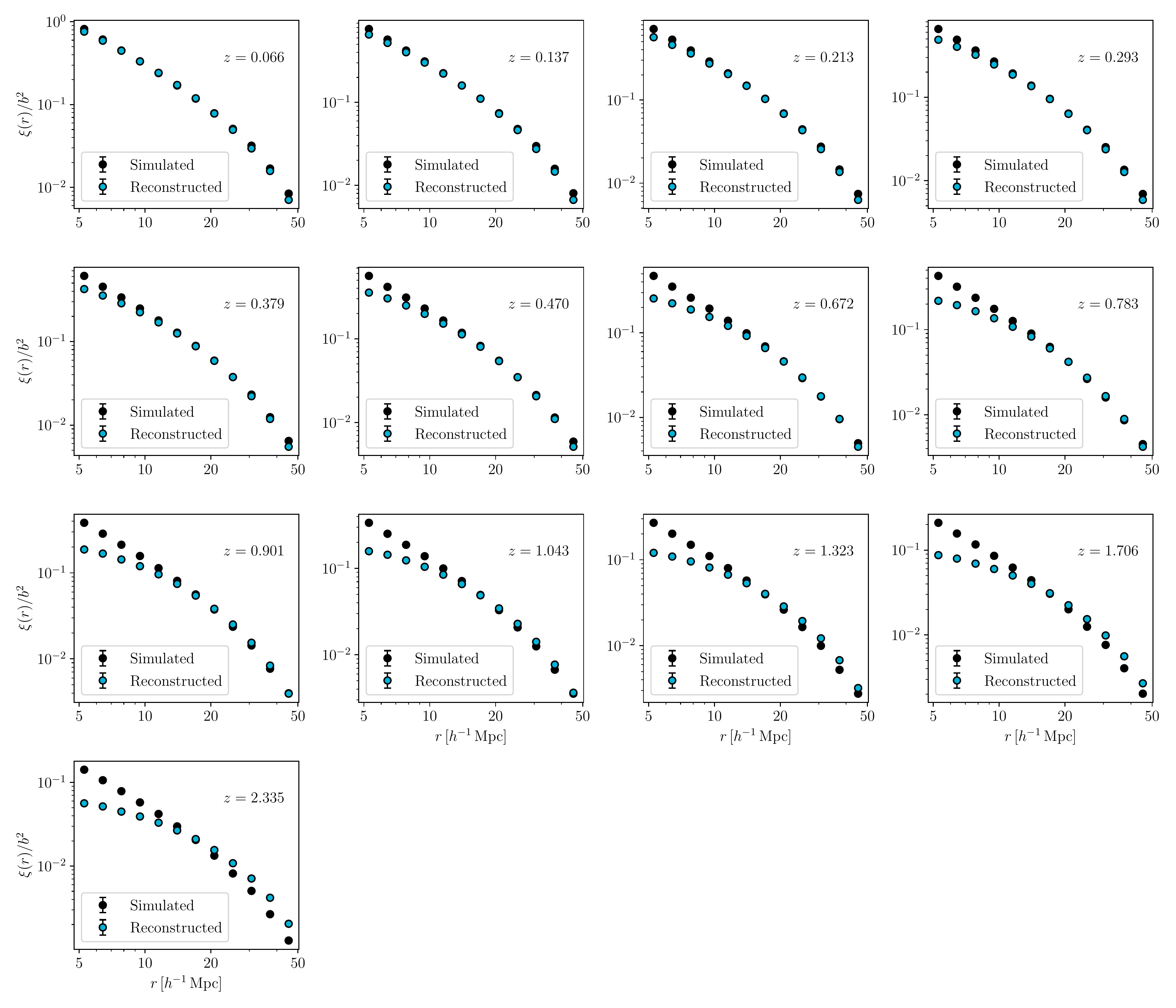} 
    \caption{Two-point correlation function, $\xi$, of simulated (black) and reconstructed (blue) galaxies, as estimated for each snapshot of the Magneticum simulation. To ease the comparison between reconstructed and simulated tracers, we show the $\xi$ normalized by the linear bias (Equation \ref{eq:fit_b}) of the considered galaxy sample. Errorbars are computed by Jaccknife resampling. 
    }\label{fig:xi_Sim_VS_Ref}
   
\end{figure*}

In this appendix we support our conclusions on the reliability of the reconstructed density field presented in Section~\ref{sec:reconCW} by comparing the two-point clustering of simulated and reconstructed tracers at different redshifts. The clustering statistics offers a global description of the reconstructed density field complementary to that of Purity (Section~\ref{subsection:Purity}) allowing us to highlight the possible biases introduced by reconstruction. 

In Figure~\ref{fig:xi_Sim_VS_Ref}, we show the two-point correlation function, $\xi$, of simulated (black) and reconstructed (blue) tracers normalised by their linear biases, $b$, as a function of redshift.
At separations larger than $10\, h^{-1}\mathrm{Mpc}$ and for $z\leq1.7$, the reconstructed and simulated $\xi/b$ are almost superimposed. Differently, for $r<10\, h^{-1}\mathrm{Mpc}$, the clustering signals progressively diverge with redshifts. 
The detected loss of power at small scales is in agreement with the decreasing Purity of clusters as a function of redshift shown in Figure~\ref{fig:Purity} and confirms the lower accuracy of eFAM at recovering the dynamics at small scales.


\bsp	
\label{lastpage}
\end{document}